\documentclass[sigconf, nonacm, pdfa]{acmart}
\setcopyright{none}
\usepackage{tabularx}
\usepackage{booktabs}
\usepackage{balance}
\usepackage{pgfplots}
\pgfplotsset{compat=1.18}
\usepackage{fancyhdr}
\usepackage[linesnumbered,vlined,ruled]{algorithm2e}
\usepackage{algpseudocode}
\usepackage{graphicx}
\usepackage{textcomp}
\usepackage[dvipsnames]{xcolor}
\usepackage{bm}
\usepackage{booktabs}
\usepackage{multirow}
\usepackage{enumitem}
\usepackage{microtype}
\usepackage{subfigure}
\usepackage{booktabs} 
\usepackage{amsfonts}

\usepackage{amssymb}
\usepackage{amsmath}
\usepackage{amsthm}
\usepackage{sansmath}
\usepackage{subfigure}
\usepackage{graphicx}
\usepackage{subcaption}
\usepackage{colortbl}
\usepackage{tikz}
\usepackage{url}
\usepackage{pifont}
\usepackage{makecell}
\usepackage{cleveref}
\usepackage{pgfplots}
\usepackage{setspace}
\usepackage{xspace}
\usepackage{listings}
\usepackage{makecell}
\usepackage{multirow}
\usepackage[dvipsnames]{xcolor}

\definecolor{LightCyan}{rgb}{0.88,1,1}
\definecolor{LightGreen}{rgb}{0.88,1,1}
\definecolor{LightRed}{rgb}{1,0.88,1}
\definecolor{LightYellow}{rgb}{1,1,0.88}
\definecolor{LightGray}{gray}{0.8}

\definecolor{Improve}{rgb}{0.88,1,1}
\definecolor{Best}{RGB}{144, 238, 144} 
\definecolor{LightRed}{rgb}{1,0.88,1}
\definecolor{SecondBest}{rgb}{1,1,0.88} 
\definecolor{Fail}{gray}{0.9}

\crefformat{section}{\S#2#1#3} 
\crefformat{subsection}{\S#2#1#3}

\long\def\comment#1{}

\usepackage[most]{tcolorbox}
\definecolor{findingbg}{RGB}{229,236,245}   
\definecolor{findingbar}{RGB}{52,92,140}    
\providecommand{\finding}{}
\renewcommand{\finding}[3]{%
  \par
  \begin{tcolorbox}[
    colback=findingbg,
    colframe=findingbar,
    boxrule=0pt,
    leftrule=2.5pt,
    arc=0pt,
    left=6pt,
    right=6pt,
    top=4pt,
    bottom=4pt]
  \phantomsection\label{finding:F#1}%
  \small\textbf{\textcolor{findingbar}{Finding #1:}}~
  \textbf{#2} #3
  \end{tcolorbox}\par}
\providecommand{\ftag}{}
\renewcommand{\ftag}[1]{\hfill{\scriptsize\hyperref[finding:#1]{\textcolor{findingbar}{[#1]}}}}

\begin{document}

\title{Cascadeability and the Realization Gap: An Empirical Study of LLM-Powered Semantic Filtering}

\newcommand{\nthesection}{\arabic{section}}

\newcounter{remark}[section]
\renewcommand{\theremark}{\nthesection.\arabic{remark}}
\newenvironment{remark}{\begin{em}
        \refstepcounter{remark}
        {\vspace{1ex}\noindent\bf Remark \theremark:}}{
        \end{em}\vspace{1ex}} 



\newcommand{\proofsketch}{\noindent{\bf Proof Sketch: }}
\newcommand{\myproof}{\noindent{\bf Proof: }}


\newcommand{\eop}{\hspace*{\fill}\mbox{$\Box$}\vspace*{1ex}}
\newcommand{\stitle}[1]{\par\noindent\textbf{#1}}
\newcommand{\etitle}[1]{\vspace{1ex}\noindent{\underline{\em #1}}}
\newcommand{\kw}[1]{{\ensuremath {\mathsf{#1}}}\xspace}

\newcommand{\algocomment}[1]{{\footnotesize $\rhd$ \emph{#1}}}

\newcommand{\mat}[2]{{\begin{tabbing}\hspace{#1}\=\+\kill #2\end{tabbing}}}
\newcommand{\m}{\hspace{0.05in}}
\newcommand{\ls}{\hspace{0.1in}}
\newcommand{\beqn}{\begin{eqnarray*}}
\newcommand{\eeqn}{\end{eqnarray*}}

\newcounter{ccc}
\newcommand{\bcc}{\setcounter{ccc}{1}\theccc.}
\newcommand{\icc}{\addtocounter{ccc}{1}\theccc.}

\newcommand{\tabincell}[2]{\begin{tabular}{@{}#1@{}}#2\end{tabular}} 

\newcommand{\kwnospace}[1]{{\ensuremath {\mathsf{#1}}}}

\newcommand{\radd}[1]{\textcolor{red}{#1}}
\newcommand{\revise}[1]{\textcolor{blue}{#1}}

\newcommand{\bigo}{\ensuremath{\mathcal{O}}}%

\newcommand{\LOTUS}{{\sl LOTUS}\xspace}
\newcommand{\BARGAIN}{{\sl BARGAIN}\xspace}
\newcommand{\ScaleDoc}{{\sl ScaleDoc}\xspace}
\newcommand{\TaskCascade}{{\sl Task Cascade}\xspace}

\newcommand{\GAT}{{\sl GAT}\xspace}
\newcommand{\SAGE}{{\sl GraphSAGE}\xspace}
\newcommand{\GCARE}{{\sl G-CARE}\xspace}
\newcommand{\WJ}{{$\texttt{WJ}$}\xspace}
\newcommand{\CS}{{$\texttt{CS}$}\xspace}
\newcommand{\CSET}{{$\texttt{CSET}$}\xspace}
\newcommand{\IMPR}{{$\texttt{IMPR}$}\xspace}
\newcommand{\JSUB}{{$\texttt{JSUB}$}\xspace}
\newcommand{\BS}{{$\texttt{BS}$}\xspace}
\newcommand{\SumRDF}{{$\texttt{SumRDF}$}\xspace}
\newcommand{\NRP}{{\sl NRP}\xspace}
\newcommand{\DeepDB}{{\sl DeepDB}\xspace}
\newcommand{\Naru}{{\sl Naru}\xspace}
\newcommand{\DBEst}{{\sl DBEst}\xspace}
\newcommand{\CONF}{{$\texttt{CON}$}\xspace}
\newcommand{\MAR}{{$\texttt{MAR}$}\xspace}
\newcommand{\ENT}{{$\texttt{ENT}$}\xspace}
\newcommand{\CTC}{{$\texttt{CTC}$}\xspace}
\newcommand{\RAND}{{$\texttt{RAN}$}\xspace}
\newcommand{\ENS}{{$\texttt{ENS}$}\xspace}
\newcommand{\GQL}{{$\texttt{GQL}$}\xspace}
\newcommand{\ORI}{{$\texttt{ORI}$}\xspace}
\newcommand{\GFlow}{{$\texttt{GFlow}$}\xspace}

\newcommand{\PubMed}{{\kw{PubMed}}\xspace}
\newcommand{\BigPatent}{{\kw{BigPatent}}\xspace}
\newcommand{\Enron}{{\kw{Enron}}\xspace}
\newcommand{\GovReport}{{\kw{GovReport}}\xspace}
\newcommand{\Court}{{\kw{Court}}\xspace}
\newcommand{\Legal}{{\kw{Legal}}\xspace}
\newcommand{\FEVER}{{\kw{FEVER}}\xspace}
\newcommand{\Movie}{{\kw{Movie}}\xspace}
\newcommand{\Steam}{{\kw{Steam}}\xspace}
\newcommand{\Wiki}{{\kw{Wiki}}\xspace}

\newcommand{\cmark}{\color{green}\ding{52}}
\newcommand{\ccross}{\color{red}\ding{56}}
\newcommand{\halfcheck}{%
    \color{blue}
  \tikz[baseline=(c.base)]{%
    \node[anchor=south west, inner sep=0pt, outer sep=0pt] (c) at (0,0) {\ding{52}};
    \begin{scope}[x={(c.south east)}, y={(c.north west)}]
      \draw[line width=1pt, line cap=round] (0.60,0.98) -- (0.95,0.50);
    \end{scope}
  }%
}

\newcommand{\loss}{\mathcal{L}}
\newcommand{\mse}{\mathcal{L}_{mse}}
\newcommand{\batch}{\mathcal{B}}
\newcommand{\relu}{\mathsf{ReLU}}
\newcommand{\softmax}{\mathsf{softmax}}
\newcommand{\COUNT}{\kw{cnt}}
\newcommand{\AVG}{\kw{avg}}
\newcommand{\SUM}{\kw{sum}}
\newcommand{\MIN}{\kw{min}}
\newcommand{\MAX}{\kw{max}}
\newcommand{\dsb}{\kw{dsb}}

\newcommand{\AGNews}{{\kw{AGNews}}\xspace}
\newcommand{\GSM}{{\kw{GSM8K}}\xspace}
\newcommand{\IMDB}{{\kw{IMDB}}\xspace}
\newcommand{\MMLU}{{\kw{MMLU}}\xspace}
\newcommand{\MRPC}{{\kw{MRPC}}\xspace}
\newcommand{\SNLI}{{\kw{SNLI}}\xspace}

\newcommand{\SST}{{\kwnospace{SST}-\kw{2}}\xspace}

\newcommand{\RoBatch}{{\textsc{RoBatch}}\xspace}

\newcommand{\RouteLLM}{{\text{RouteLLM}}\xspace}
\newcommand{\FrugalGPT}{\text{FrugalGPT}\xspace}
\newcommand{\BATCHER}{{\text{BATCHER}}\xspace}
\newcommand{\OBP}{{\text{OBP}}\xspace}

\newcommand{\tbd}{{\color{green}TBD}}

\newcommand{\Model}{\mathcal{M}}
\newcommand{\Dataset}{\mathcal{Q}}
\newcommand{\Prob}{\mathcal{P}}

\newcommand{\Real}{\mathbb{R}}

\newcommand{\norm}[1]{\left\lVert#1\right\rVert}


\newcommand{\score}{\ensuremath{\text{score}}\xspace}
\newcommand{\SelectFromBatch}{\textit{SelectFromBatch}\xspace}


\author{Nan Hou}
\affiliation{%
  \institution{The Chinese University of Hong Kong}
  \country{}
}
\email{nhou@se.cuhk.edu.hk}

\author{Kangfei Zhao}
\affiliation{%
  \institution{Beijing Institute of Technology}
  \country{}
}
\email{zkf1105@gmail.com}

\author{Jiadong Xie}
\affiliation{%
  \institution{The Chinese University of Hong Kong}
  \country{}
}
\email{jdxie@se.cuhk.edu.hk}

\author{Jeffrey Xu Yu}
\affiliation{%
  \institution{HKUST (Guangzhou)}
  \country{}
}
\email{jeffreyxuyu@hkust-gz.edu.cn}

\begin{abstract}

Semantic filters enable natural language predicates over unstructured documents, but evaluating every document with a large language model (LLM) oracle is expensive. Proxy cascades reduce oracle invocations using cheaper surrogates, but oracle reduction alone cannot distinguish proxy limitations from calibration bottlenecks and establish practical efficiency. We present a unified benchmark of four representative systems, which formulate query processing as a three-stage workflow comprising proxy construction, threshold calibration, and cascade inference, using a two-threshold acceptance rule. We introduce cascadeability: the maximum empirical oracle reduction attainable from fixed proxy scores at a specified accuracy target, computed retrospectively using complete oracle labels. We investigate 8 research questions, spanning proxy quality, end-to-end performance, workload sensitivities, and component ablations, and reveal our key findings accordingly. Comparing cascadeability with achieved oracle reduction yields a realization gap that, interpreted alongside accuracy, exposes distinct proxy and calibration bottlenecks. The benchmark and diagnostics support evaluation beyond oracle reduction and motivate joint proxy–calibration design, class-sensitive quality assessment, and end-to-end cost optimization. We release a dataset-query benchmark, containing 10 real-world workloads across various domains and difficulties, with annotations of two LLM families. 

\end{abstract}

\maketitle

\noindent\textbf{Code and Data Availability.}
The source code and benchmark datasets are
available at
\url{https://github.com/Anto-an/semantic_filter_benchmark}.

\section{Introduction}

Semantic filters extend classical database selection to natural language predicates over textual attributes or unstructured documents, allowing users to express semantic conditions that are difficult to express using symbolic selection operators~\cite{DBLP:journals/pvldb/lotus, shankar2025docetl}.
A reference physical execution evaluates the predicate on every document using a large language model (LLM), incurring substantial query latency and monetary cost. 
To reduce this overhead, semantic filter systems have proposed proxy cascades that employ cheaper surrogate models or proxy tasks to pre-filter documents, and direct unresolved documents to the reference LLM~\cite{DBLP:journals/pvldb/lotus, DBLP:journals/pacmmod/bargain, DBLP:journals/corr/taskcascades, DBLP:journals/corr/scaledoc}. 
The design objective is not simply to replace the expensive LLM with cheaper models,  but to reduce execution cost under a specified accuracy quality target. The accuracy target is typically measured as the agreement with the reference LLM that serves as an oracle.

Evaluating whether a cascade achieves this objective requires distinguishing the opportunity available to the system from its ability to exploit that opportunity. 
The oracle reduction rate (ORR) measures the fraction of documents bypassed by the proxy, directly determining the cost savings of oracle invocations.
Systems generally favor high ORR under the quality target, but ORR alone cannot distinguish proxy limitations from calibration bottlenecks or establish practical efficiency.
Precisely, low ORR can arise for two different reasons: the proxy itself may provide little opportunity to bypass the oracle invocations while preserving the required accuracy target, or the proxy scores may support substantial reduction, but the system fails to calibrate effective proxy thresholds to realize the reduction. 
Furthermore, a high ORR does not necessarily indicate an effective cascade, which may accompany violations of the accuracy target. In addition, the costs of proxy construction, proxy scoring, and threshold calibration are not negligible and may offset the savings from the reduction of oracle calls.
Therefore, understanding cascade performance requires connecting available opportunity, realized reduction, and practical efficiency.

Existing systems make different design choices at each stage of proxy construction and proxy cascade. 
For example, \LOTUS~\cite{DBLP:journals/pvldb/lotus} and \BARGAIN~\cite{DBLP:journals/pacmmod/bargain} adopt small LLMs as surrogate models but differ in threshold calibration and quality estimation. 
\ScaleDoc~\cite{DBLP:journals/corr/scaledoc} trains a lightweight surrogate on document embeddings, while \TaskCascade~\cite{DBLP:journals/corr/taskcascades} constructs a multi-stage cascade of task-level proxies by agentic greedy search. These differences make it difficult to attribute observed performance to the surrogate alone or to the procedure. 
Although prior studies have examined several aspects of quality and efficiency, the relationship between proxy opportunity and its realization remains insufficiently characterized across these designs under comparable conditions. We investigate this relationship through 8 research questions covering proxy discrimination and confidence, attainable and realized oracle reduction, target satisfaction and execution cost, workload sensitivity, and cross-framework component combinations. Together, these questions move the evaluation beyond identifying which system performs best to understanding where its gains and limitations arise.

We present a unified benchmark and experimental study that organizes the four representative systems into three phases: proxy construction, threshold calibration, and cascade inference. We evaluate them under an accuracy target, with explicit adaptations where their native formulations differ. 
The benchmarking is conducted on 10 dataset–predicate workloads spreading the original testing of the four systems with two LLM families. 
We propose a retrospective diagnostic, cascadeability $\kappa_{\alpha}$, for proxy cascades under the two-threshold acceptance rule,  which is the maximum oracle reduction attainable from a fixed proxy score distribution at a specified accuracy target ${\alpha}$. Comparing this quantity with achieved ORR yields a realization gap, which we interpret together with achieved accuracy to distinguish limited proxy opportunity from incomplete exploitation. 
We complement this analysis with target satisfaction measurements, token costs, end-to-end latency, and component-switch ablations.

Across 8 research questions, our benchmarking reveals complementary findings spanning proxy behavior and confidence, cascadeability and realization gap, reliability, efficiency, workload sensitivity, and cross-component ablations, where limitations exist in both cascadeability and realization across all the systems.
Altogether, these findings move beyond system rankings to identify distinct barriers to reliable and efficient semantic filtering. In a nutshell, semantic filter systems favor joint proxy–calibration design, workload-aware and class-sensitive quality control, and end-to-end cost optimization.
Finally, we outline 5 future research directions that broaden the scope of semantic filter processing.
Within the current scope of single-predicate filters over static document collections, the contributions of this paper are summarized as follows:
\begin{itemize}[leftmargin=*]
    \item \textbf{An empirical characterization of cascade bottlenecks and trade-offs.} We distinguish limitations in proxy opportunity from limitations in its realization, and connect these observations to confidence behavior, target satisfaction, token cost, and latency. 
    \item \textbf{A common evaluation framework with retrospective diagnostics.} We place four representative systems within a unified framework and evaluate them by a common protocol, making their proxy-construction and calibration choices explicit. 
    \item \textbf{A comprehensive semantic filter test suite.} We collect and release a dataset-predicate test suite containing 10 real-world workloads with annotations of two LLM families. Based on these datasets, we also construct specific workloads with controlled query selectivities and oracle confidence score distributions. 
\end{itemize}

\section{Preliminaries}
\stitle{Semantic Filters.}
Let $D=\{x_1,\dots,x_n\}$ be a collection of $n$ documents and $q$ a natural-language predicate, such as ``the opinion reverses a lower-court ruling''. A semantic filter returns the documents in $D$ that satisfy $q$. 
The \emph{reference execution} of a semantic filter invokes a capable yet expensive oracle model, i.e., a large language model (LLM), $M_{o}$ on every document, and the obtained reference label of a document $x_i$ is denoted as $y_i = M_o(x_i, q) \in \{0, 1\}$, where $y_i = 1$ indicates that $M_o$ judges $x_i$ to satisfy $q$, and $y_i = 0$ otherwise. Thereby, the reference execution invokes $M_{o}$ $n$ times to process the collection $D$, which is expensive w.r.t. both token cost and latency. 

\stitle{Proxy Cascade.} Semantic filter systems reduce the cost of reference execution with a \emph{proxy cascade} paradigm~\cite{DBLP:journals/pvldb/lotus, DBLP:journals/corr/taskcascades, DBLP:journals/corr/scaledoc,DBLP:journals/pacmmod/bargain}. 
These systems employ an inexpensive surrogate model $M_s$ that produces a proxy score $\widehat{s_i} = M_s(x_i, q) \in [0, 1]$ for each document $x_i \in D$, indicating how likely $x_i$ satisfies $q$. 
Throughout this paper, a proxy stage uses a \emph{two-threshold} acceptance rule to process the documents $D$. The two thresholds $\tau^-, \tau^+ \in [0, 1]$ with $\tau^- \leq \tau^+$ are calibrated from a small sample of documents $D' \subset D$ whose oracle labels are collected by invoking the oracle model $M_o$.
Proxy cascade, therefore, processes remaining documents $D \setminus D'$ in a two-stage paradigm: 
\begin{itemize}[leftmargin=*]
    \item Surrogate invocation. If $M_s(x_i, q) \leq \tau^-$ or $M_s(x_i, q) \geq \tau^+$, the system immediately returns False or True, respectively. 
    \item Oracle fallback. If $ M_s(x_i, q) \in (\tau^-, \tau^+)$, the system resorts to the oracle model $M_o$ and returns the result $M_o(x_i, q)$.
\end{itemize} 
On the calibration set $D'$, the two thresholds $\tau^-, \tau^+$  are calibrated according to the expectation of the quality target: 
\begin{align}
    \mathbb{P}\left[
        \frac{1}{n}\sum_{i=1}^{n}
        \mathbb{I}(
            \widehat{y_i} = y_i
        )
        \geq \alpha
    \right]
    \geq 1-\delta,
    \label{eq:quality-target}
\end{align}
where $\widehat{y_i}$ is the final output of the cascade, $\alpha$ is the accuracy target and $\delta$ is the failure probability. The systems support different quality targets such as accuracy, precision, recall, and F1. For a common evaluation, we use the accuracy target in Eq.~\eqref{eq:quality-target} as the quality target in this benchmark because it incurs minimal changes to the systems being studied.

\section{A Unified Proxy Cascade Framework}

\begin{figure}
    \centering
    \includegraphics[width=\linewidth]{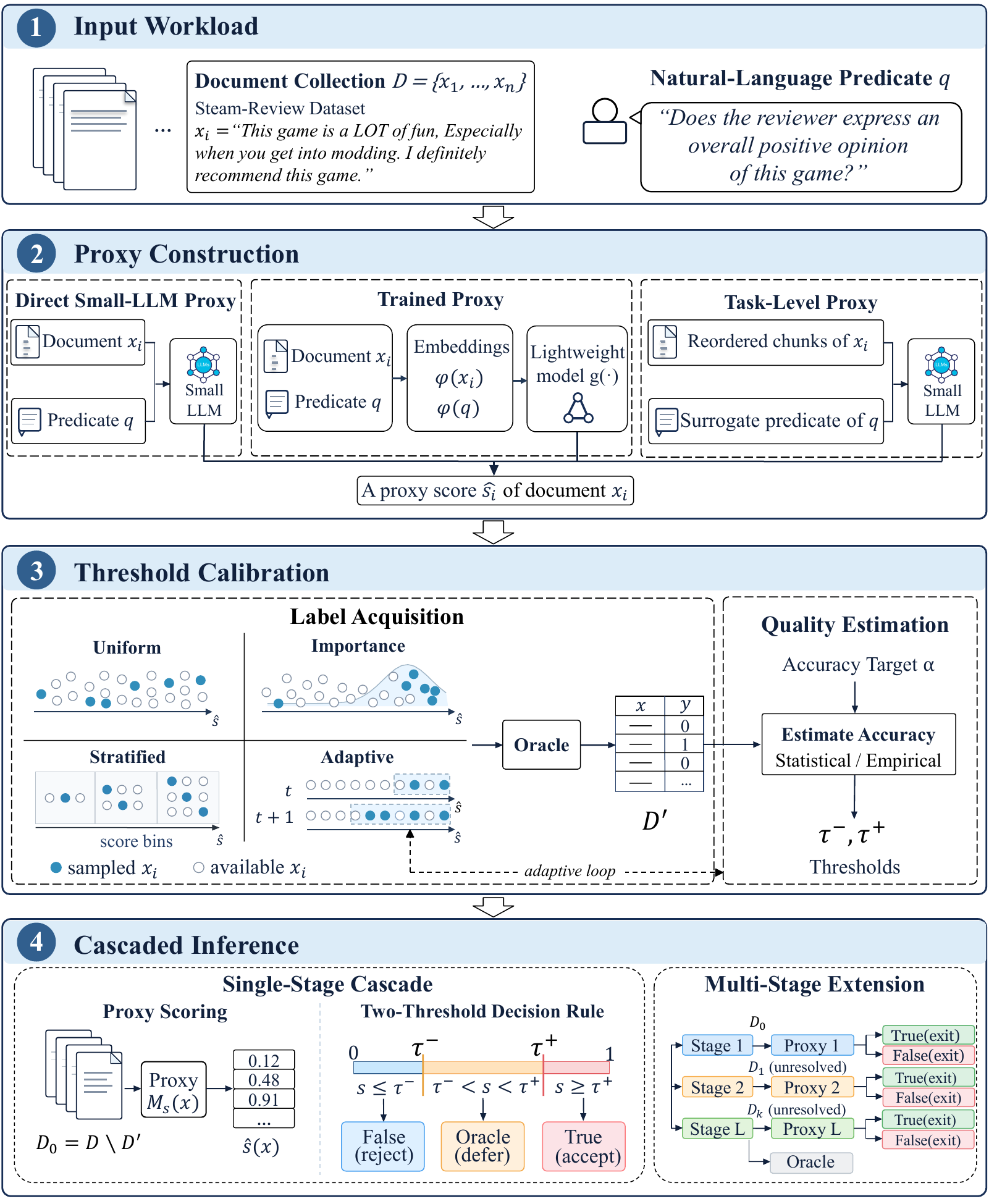}
    \caption{Cascade framework of semantic filter}
    \label{fig:pipline}
\end{figure}

We unify the deployment of proxy cascades in the systems being studied into a general framework for a holistic review, although there are differences in specific implementations. As illustrated in Fig.~\ref{fig:pipline}, a workload consists of a document collection $D$ and a natural-language predicate $q$. The framework comprises three phases: proxy construction, threshold calibration, and cascade inference.


\stitle{Proxy Construction.}
This phase constructs one or more inexpensive proxies that assign a score $\hat{s}_i$ to each document $x_i$. Fig.~\ref{fig:pipline} illustrates three representation designs: direct small LLM, trained lightweight, and task-level proxies. 
System instantiations differ along five dimensions: 
\emph{Surrogate model} is either a smaller LLM or a lightweight model trained per predicate on precomputed document embeddings; 
\emph{Proxy score} is derived from the surrogate model and is either a token probability, the confidence of the predicted label, or a similarity between embeddings, indicating the likelihood that the document passes the predicate;
\emph{Task} is either an original predicate or a surrogate task that is easier than the oracle, such as a prerequisite predicate;
\emph{Input} is either the original document, or pruned or reordered chunks or the embedding of the document computed offline;
\emph{Supervision} involves zero-shot inference, few-shot prompting, or supervised training using oracle-labeled examples.

\stitle{Threshold Calibration.} Given the proxies, this phase selects two thresholds $\tau^-, \tau^+$  from a calibration set $D' = \{x_1, \cdots\}$ where each document $x_i$ is annotated with its oracle label $y_i$, following the two-threshold acceptance rule and the default accuracy target in Eq.~\eqref{eq:quality-target}. 
This phase comprises two intertwined design dimensions.
\emph{Label acquisition} decides which and how many documents to label as the calibration set $D' \subseteq D$, obtaining corresponding oracle labels $y_i$. Systems employ uniform, importance, stratified, or adaptive sampling, potentially guided by proxy scores. Sampling may be performed in a single round or iteratively, with subsequent selections informed by previous observations and quality estimates.
\emph{Quality estimation} decides how to evaluate the expected accuracy and quantify the uncertainty, using empirical estimates or statistical methods such as the Central Limit Theorem, a betting-based test, or Bernstein bound. These estimates guide the selection of a pair of thresholds $\tau^-$ and $\tau^+$ for each proxy stage.
In multi-stage systems, the phases of proxy construction and threshold calibration may be interleaved, and multi-stage proxies are jointly optimized. 

\stitle{Cascaded Inference.} With the cascaded proxy and the thresholds, the framework processes each document $x_i \in D \setminus D'$ using the two-stage workflow of surrogate invocation and oracle fallback. Multi-stage cascade extends the two-stage workflow, where each stage processes only the documents that reach the stage.

\section{Systems Under Study}
\label{sec:systems}

\begin{table*}[t]
    \centering
    \footnotesize
    \caption{Comparison of proxy construction and threshold calibration methods}
    \label{tab:proxy-threshold-comparison}
        \begin{tabular}{@{}llllllll@{}}
            \toprule
            & \multicolumn{5}{c}{\textbf{Proxy Construction}}
            & \multicolumn{2}{c}{\textbf{Threshold Calibration}} \\
            \cmidrule(lr){2-6}
            \cmidrule(l){7-8}
            & \textbf{Surrogate model}
            & \textbf{Proxy Score}
            & \textbf{Task}
            & \textbf{Input}
            & \textbf{Supervision}
            & \textbf{Label acquisition}
            & \textbf{Quality estimation} \\
            \midrule
            \LOTUS~\cite{DBLP:journals/pvldb/lotus}
            & small LLM
            & \makecell[l]{Quantile-normalized \\ predictive probability}
            & original predicate
            & original document
            & zero-shot
            & \makecell[l]{Importance  \& \\ Uniform sample}
            & CLT bound \\ \midrule

            \BARGAIN~\cite{DBLP:journals/pacmmod/bargain}
            & small LLM
            & predictive probability
            & original predicate
            & original document
            & zero-shot
            & \makecell[l]{Target-aware \\ adaptive sample}
            & Betting-based test \\ \midrule

            \ScaleDoc~\cite{DBLP:journals/corr/scaledoc}
            &  \makecell[l]{multilayer \\ perceptron}
            & \makecell[l]{Cosine similarity \\ of embeddings}
            & \makecell[l]{original predicate}
            & document embedding
            & \makecell[l]{collected \\ training data}
            & Stratified sample
            & Bernstein bound \\ \midrule
            \TaskCascade~\cite{DBLP:journals/corr/taskcascades}
            & small LLM
            & predictive probability
            & \makecell[l]{original or agent-\\generated predicate}
            & \makecell[l]{prefix of reordered \\ document chunks}
            & few-shot
            & Uniform sample
            & Empirical \\ 
            \bottomrule
        \end{tabular}%
\end{table*}

Table~\ref{tab:proxy-threshold-comparison} summarizes how each system instantiates the proxy cascade framework. For each system being studied, we describe its native design and our configuration for the common testing protocol (\cref{sec:systems:design}), and elaborate on their quality estimation and control (\cref{sec:systems:quality}).

\subsection{Native Design \& Our Configuration}
\label{sec:systems:design}

\etitle{\LOTUS~\cite{DBLP:journals/pvldb/lotus}}. 
\LOTUS employs a pretrained smaller LLM as a surrogate model that evaluates the predicate $q$ on every document in $D$. The proxy scores are the log-probabilities of the \texttt{True}/\texttt{False} output tokens, normalized to the index of the quantile bucket over the whole collection (50 buckets by default). 
Following SUPG~\cite{DBLP:journals/pvldb/KangGBHZ20}, \LOTUS draws an importance sample with replacement. Sampling probabilities are proportional to the square root of the proxy score and are defensively mixed with uniform sampling. \LOTUS labels the sample with the oracle.
Using the central limit theorem and the normal approximation on the samples, \LOTUS selects $\tau^+$ and $\tau^-$ to reach user-specified Precision and Recall targets, respectively, by allotting a failure probability $\delta/2$. 
The native implementation of \LOTUS does not support an accuracy target. For unified benchmarking, we extend  \LOTUS with an accuracy target by keeping its proxy scores, sampling design, and normal-approximation bounds, and replacing the estimated metric to select $\tau^-,\tau^+$. 

\etitle{\BARGAIN~\cite{DBLP:journals/pacmmod/bargain}.} Similar to \LOTUS, \BARGAIN evaluates $q$ on each document with a smaller LLM. However, it has different variants that support accuracy, precision, and recall targets under finite-sample guarantees. We use the accuracy target variant in this paper.
Taking the predictive probabilities of the output \texttt{True} Token directly as the proxy scores, \BARGAIN partitions all documents into score regions defined by pre-specified intervals (e.g., 20 score subranges), where each region identifies a candidate threshold. \BARGAIN starts sampling from the region with the highest proxy scores and progressively expands to lower-score regions. Within each region, it samples tuples and invokes the oracle LLM to test whether the tuples in that region can meet a user-specified quality target under the Betting-based test~\cite{jiang2024estimating}. The region-wise procedure terminates once \BARGAIN identifies a minimum threshold $\tau$ such that including any lower-score region would violate the target quality. The native implementation of \BARGAIN is a single threshold cascade. Here, we adapt it to a two-threshold cascade by renormalizing the proxy's {True}/{False} token probabilities and setting $\tau^- = 1-\tau$, $\tau^+ =\tau$. 

\etitle{\ScaleDoc~\cite{DBLP:journals/corr/scaledoc}.}
\ScaleDoc uses a multilayer perceptron (MLP) as the surrogate, which is trained per query on document embeddings. Given uniformly sampled training data annotated by the oracle model, the MLP is trained by contrastive learning that distinguishes the positive and negative samples and enforces bipolarity in the latent space. The cosine similarity between the resulting document embedding and the predicate embedding serves as the proxy score. 
For threshold calibration, \ScaleDoc first partitions score ranges into bins, then draws a separate calibration sample by stratified sampling with allocation proportional to bin populations, and labels the samples by oracle invocation. From the labeled sample, it reconstructs the score density by linear interpolation and moving average smoothing.
For candidate threshold pairs on the bin boundaries, \ScaleDoc estimates the proxy quality and selects the thresholds with the lowest oracle deferral subject to the accuracy target. 

\etitle{\TaskCascade~\cite{DBLP:journals/corr/taskcascades}.}
\TaskCascade generalizes the two-stage proxy to a multi-stage proxy, where each stage is a task-level proxy, specified by a surrogate LLM, a surrogate predicate, and a fraction of document chunks. 
First, each document is restructured so that the
content most relevant to the original predicate $q$ appears first.
Specifically, \TaskCascade reorders the chunks of each document by their relevance to $q$, and a stage (task) reads only a prefix of the chunks of its subsequent stage, with full document fallback to the oracle, which enables prefix cache reuse. 
Guided by a development set, an LLM agent then iteratively adds surrogate predicates, i.e., simpler, decomposed, or otherwise correlated predicates. 
\TaskCascade sets its thresholds empirically on
the development sample, without a held-out calibration step. For each candidate task, each class threshold is set to the lowest value at which at least an $\alpha$ fraction of the predictions accepted for that class agree with the oracle. 
Candidate tasks are assembled greedily w.r.t. minimizing token cost. At inference time, a
document exits at the first stage whose proxy score exceeds the threshold.

\subsection{Quality Estimation \& Control}
\label{sec:systems:quality}

Let $\boldsymbol{\tau}=(\tau^-,\tau^+)$ denote the threshold configuration for a single proxy stage.
We denote by $\Gamma_D(\boldsymbol{\tau})$ the fraction of documents resolved by the proxy, i.e., the oracle reduction rate (ORR), and by $Q_D(\boldsymbol{\tau})$ the accuracy of these proxy-resolved predictions.
Since deferred documents are processed by the oracle, the overall accuracy is $\mathrm{Acc}_D(\boldsymbol{\tau})=1-\Gamma_D(\boldsymbol{\tau})(1-Q_D(\boldsymbol{\tau})).$
Threshold calibration estimates this quality using oracle-labeled samples and selects thresholds to minimize oracle deferral subject to $\mathrm{Acc}_D(\boldsymbol{\tau})\ge\alpha$.

\etitle{\LOTUS~\cite{DBLP:journals/pvldb/lotus}}
uses importance sampling mixed with uniform sampling and CLT-based normal approximation bounds.
Let $\{(x_{i_j},y_{i_j})\}_{j=1}^{m}$ denote the $m$ oracle-labeled calibration draws, where $x_{i_j}$ is sampled with replacement from $D$ with probability $p_{i_j}$.
For a threshold configuration $\boldsymbol{\tau}$, let
$Z_j(\boldsymbol{\tau})={\mathbb{I}[\widehat{y}_{i_j}(\boldsymbol{\tau})\neq y_{i_j}]}/{n /p_{i_j}}$ be the importance-weighted error indicator, where $\widehat{y}_{i_j}(\boldsymbol{\tau})$ is the final cascade prediction.
Let $\overline Z(\boldsymbol{\tau})$ and $\widehat{\sigma}_Z(\boldsymbol{\tau})$ denote the sample mean and standard deviation of the $m$ weighted error indicators, respectively.
The lower accuracy bound is $L_{\mathrm{CLT}}(\boldsymbol{\tau})=1-\overline Z(\boldsymbol{\tau})-z_{1-\delta}\frac{\widehat{\sigma}_Z(\boldsymbol{\tau})}{\sqrt{m}},$ where $z_{1-\delta}$ is the $(1-\delta)$-quantile of the standard normal distribution.

\etitle{\BARGAIN~\cite{DBLP:journals/pacmmod/bargain}} uses adaptive sampling and an anytime-valid betting-based test to certify proxy accuracy.
For a threshold configuration $\boldsymbol{\tau}$, let $R(\boldsymbol{\tau})\subseteq D$ be the set of documents resolved by the proxy, with $N=|R(\boldsymbol{\tau})|$.
\BARGAIN samples documents from this set without replacement.
Let $X_j=\mathbb{I}[\widehat{y}_{i_j}=y_{i_j}]$ indicate whether the proxy prediction agrees with the oracle on the $j$-th sampled document.
To test whether the accepted-region accuracy $Q_D(\boldsymbol{\tau})$ exceeds $\alpha$, \BARGAIN constructs a betting process of the form $K_t=\prod_{j=1}^{t}(1+\lambda_j(X_j-\mu_j))$, $\mu_j=\frac{N\alpha-\sum_{h=1}^{j-1}X_h}{N-j+1}$, where $t$ is the number of observed labels, $\mu_j$ accounts for sampling without replacement under the null hypothesis $Q_D(\boldsymbol{\tau})\le\alpha$, and $\lambda_j$ is a nonnegative betting fraction chosen from preceding observations and constrained to keep the process nonnegative. A threshold is certified when $K_t\ge 1/\delta$. Under the null hypothesis, the betting process satisfies $\mathbb{P}[\sup_t K_t\ge 1/\delta]\le\delta$.
Since $\mathrm{Acc}_D(\boldsymbol{\tau})=1-\Gamma_D(\boldsymbol{\tau})(1-Q_D(\boldsymbol{\tau}))$, certifying $Q_D(\boldsymbol{\tau})\ge\alpha$ also guarantees the overall accuracy target.

\etitle{\ScaleDoc~\cite{DBLP:journals/corr/scaledoc}} uses stratified sampling to obtain oracle-labeled calibration data and reconstructs the positive and negative score distributions through interpolation and smoothing.
Let $\widetilde F^+(u)$ and $\widetilde F^-(u)$ denote the estimated fractions of oracle-positive and oracle-negative documents, respectively, with proxy scores at most $u$.
For $\boldsymbol{\tau}=(\tau^-,\tau^+)$, its accuracy adaptation estimates $\widehat{\mathrm{Acc}}_{\mathrm{SD}}(\boldsymbol{\tau})=1-\widetilde F^+(\tau^-)-\bigl(\widetilde F^-(1)-\widetilde F^-(\tau^+)\bigr),$ where the two subtracted terms estimate false negatives and false positives resolved by the proxy. \ScaleDoc selects the threshold pair that minimizes oracle deferral while meeting the estimated accuracy target.
Its original theoretical analysis derives a Bernstein-based margin for an F1 target. Writing $\beta$ for this original target, the margin is $\epsilon_{\mathrm{SD}}=\left(\sigma_W+(1-\beta)\sigma_P\right)\sqrt{\frac{4\ln(4/\delta)}{m}}+\frac{(8-6\beta)\ln(4/\delta)}{3m},$ where $m$ is the calibration sample size, $\sigma_W$ is the standard deviation of the weighted error indicator $W=(1-\beta/2)\mathbb{I}[y_i=1,\widehat{s_i}<\tau^-]+(\beta/2)\mathbb{I}[y_i=0,\widehat{s_i}>\tau^+]$, and $\sigma_P$ is the standard deviation of $\mathbb{I}[y_i=1]$.

\etitle{\TaskCascade~\cite{DBLP:journals/corr/taskcascades}} uses empirical accuracy on an oracle-labeled development set $D_{\mathrm{dev}}$ to determine class-specific thresholds at each proxy stage.
For stage $\ell$ and predicted class $c\in\{0,1\}$, let $S_{\ell,c}(\tau)$ be the subset of development documents predicted as class $c$ with confidence at least $\tau$.
The empirical accuracy is $\widehat Q_{\ell,c}(\tau)=\frac{1}{|S_{\ell,c}(\tau)|} \sum_{x_i\in S_{\ell,c}(\tau)}\mathbb{I}[y_i=c].$
\TaskCascade selects the smallest threshold $\tau$ for which $S_{\ell,c}(\tau)$ is nonempty and $\widehat Q_{\ell,c}(\tau)\ge\alpha$, where $\alpha$ is the target accuracy. Candidate stages are then assembled greedily to minimize inference cost. Unlike uncertainty-aware statistical bounds, this empirical criterion includes no uncertainty margin and provides no finite-sample guarantee in the base variant.

\section{Evaluation Framework}
\label{sec:cascadeability}

\begin{table*}[t]
\footnotesize
\centering
\caption{Eight research questions in our evaluation framework.}
\label{tab:rq_methods}
\begin{tabular}{llcccc}
\toprule
Phase & Research Question & \LOTUS & \BARGAIN & \ScaleDoc & \TaskCascade \\
\midrule

\multirow{3}{*}{Proxy Construction}
    & \ding{182} How well do proxy scores distinguish positive and negative documents?
    & \ccross & \halfcheck & \halfcheck & \ccross \\ \cmidrule{2-6}
    & \ding{183} Does proxy deliver confidence correspond to practical accuracy?
    & \ccross & \cmark & \halfcheck & \ccross \\ \cmidrule{2-6}

    &\multirow{2}{*}{\makecell[l]{\ding{184} How much oracle-reduction opportunity does a proxy provide, \\ and how does the two-threshold cascade realize?}}
    & \multirow{2}{*}{\ccross}
    & \multirow{2}{*}{\halfcheck}
    & \multirow{2}{*}{\halfcheck}
    & \multirow{2}{*}{\ccross} \\ 

\cmidrule(lr){1-1}

\multirow{6}{*}{Proxy Cascade}
    & & & & & \\ \cmidrule{2-6}

    & \ding{185}  What Accuracy, ORR, and TSR does each system achieve?
    & \cmark & \cmark & \cmark & \cmark \\ \cmidrule{2-6}

    & \ding{186}  What is the token cost and end-to-end execution latency?
    & \cmark & \halfcheck & \cmark & \cmark \\ \cmidrule{2-6}

    & \ding{187} How does performance vary with query selectivity?
    & \ccross & \cmark & \cmark & \ccross \\ \cmidrule{2-6}

    & \ding{188} How does performance vary across different oracle confidence score distributions?
    & \ccross & \halfcheck & \halfcheck & \ccross \\ \cmidrule{2-6}

    & \ding{189} What if we replace surrogate models or calibration procedures across systems?
    & \halfcheck & \cmark & \cmark & \halfcheck \\

\bottomrule
\end{tabular}
\end{table*}


We organize our study around eight research questions (RQ), spanning proxy quality, end-to-end performance, robustness, and component ablations.  
Table~\ref{tab:rq_methods} summarizes these questions and their coverage in the original system studies. 
Here, RQ \ding{182} and RQ \ding{183} study the prediction quality and the beliefs of the surrogate model, respectively. 
RQ \ding{184} studies the upper limit of the oracle reduction rate of a proxy, which we define as a new posterior indicator, \emph{cascadeability}, that separates the opportunity supplied by a fixed proxy task from the ability of a practical system to exploit it.
RQ \ding{185} and RQ \ding{186} study the overall effectiveness and efficiency of the four systems, respectively.
In addition, RQ \ding{187} and RQ \ding{188} study the robustness of the four systems under two types of controlled variations of data distribution.
Finally, RQ \ding{189} ablates the effect of cross-system components to study the interplay of surrogate model and score calibration.
We use {\cmark} to denote that the question is directly evaluated, {\halfcheck} denotes that a similar question is explored, but the metrics, control variables, and factors are distinct, while {\ccross} denotes that the question is under-explored. 
The following subsections define the cascadeability diagnostic (\cref{sec:framework:cascadeability}) and the common evaluation protocol~(\cref{sec:framework:protocol}).

\subsection{Cascadeability \& the Realization Gap}
\label{sec:framework:cascadeability}



Oracle reduction rate (ORR) alone does not reveal why a system invokes the oracle frequently. Given an accuracy target, a system can suffer from low ORR because the proxy score distribution does not permit reliable decision boundaries, or because practical calibration fails to exploit available bypassing opportunities. 
For a fixed proxy cascade with $L$ proxy stages, we separate these cases with cascadeability $\kappa_\alpha$, the maximum oracle reduction attainable at the accuracy target $\alpha$. It is defined as the highest empirical ORR $\Gamma_D(\boldsymbol{\tau})$ achievable over all valid threshold configurations. The threshold configuration is represented as $\boldsymbol{\tau} = ((\tau^-_1, \tau^+_1), \ldots, (\tau^-_L, \tau^+_L))$:
\begin{align}
    \kappa_\alpha=\max_{\boldsymbol{\tau}}\Gamma_D(\boldsymbol{\tau})\quad\text{s.t.}\quad\mathrm{Acc}_{D}(\boldsymbol{\tau})\ge\alpha. \label{eq:cascadablity}
\end{align}
Together, $\kappa_\alpha$ and achieved ORR characterize
the oracle-reduction opportunity provided by the proxy
and the extent to which the system's threshold calibration
realizes this opportunity. For a single-stage cascade, we compute $\kappa_\alpha$ by searching over the sorted proxy scores using binary search to identify the optimal threshold pair under the accuracy constraint.  For multi-stage cascades, we prune the joint threshold search space and decompose the remaining optimization into smaller mixed-integer linear programming (MILP) subproblems, which we solve to optimality using HiGHS~\cite{huangfu2018parallelizing}.
Since this computation requires complete oracle labels on $D$, $\kappa_\alpha$ serves as a retrospective diagnostic rather than a practical calibration method, and does not depend on the failure probability $\delta$.


For a practical cascade with threshold configuration
$\boldsymbol{\tau}$, we define its realization
gap as $\kappa_\alpha-\Gamma_D(\boldsymbol{\tau})$.
For runs satisfying the accuracy target, a smaller gap
indicates that more of the available oracle-reduction
opportunity is realized. A negative gap necessarily
indicates a violation of the accuracy target. Varying $\alpha$ yields the cascadeability frontier
$\alpha\mapsto\kappa_\alpha$, which is monotonically
non-increasing as stricter accuracy targets reduce
the feasible set of threshold configurations.

\subsection{Evaluation Protocol}
\label{sec:framework:protocol}

\stitle{Model Settings.}
We use two LLM families, Qwen and Gemma, wherein Qwen 3.6 27B~\cite{qwen3.6-27b} and Gemma 4 12B~\cite{gemmateam2026gemma4} are used as the oracle models.
The surrogate models of \LOTUS, \BARGAIN, and \TaskCascade are the smaller LLMs in these two families, Qwen 3.5 9B~\cite{qwen3.5} and Gemma 4 Effective 4B (E4B), respectively. Their input and output token monetary costs (\$ per million tokens) are as follows:


\begin{table}[h]

{\footnotesize
\begin{tabular}{l|r r r r}
\toprule
Price (\$) & Qwen 3.6 27B & Qwen 3.5 9B & Gemma 4 12B & Gemma 4 E4B \\
\midrule
Input & 0.29 & 0.08 & 0.10 & 0.02 \\
Output & 2.50 & 0.13 & 0.30 & 0.10 \\
\bottomrule
\end{tabular}
}
\end{table}

\begin{table*}[t]
    \centering
    \caption{
        Profiles of 10 dataset-predicate workloads
    }
    \label{tab:dataset-statistics}
    \scriptsize
    \setlength{\tabcolsep}{2.8pt}

    \begin{tabular}{llrrccccccc l }
        \toprule
        \multirow{2}{*}{Dataset} & \multirow{2}{*}{Domain} & \multirow{2}{*}{$|D|$} & \multirow{2}{*}{\makecell[c]{Avg. \\ Words}} & \multirow{2}{*}{\makecell[c]{Predicate \\ Difficulty}} & \multicolumn{2}{c}{ Selectivity}  & \multirow{2}{*}{\LOTUS}
        & \multirow{2}{*}{\BARGAIN} 
        & \multirow{2}{*}{\ScaleDoc} 
        & \multirow{2}{*}{\TaskCascade} 
        & \multirow{2}{*}{Predicate summary (\texttt{True} if \ldots)}  \\ \cmidrule{6-7}
         & & & & & Gemma 4 12B & Qwen 3.6 27B &
        & 
        &
        &
        &  \\
        \midrule

        \PubMed
        & Biomedical
        & 10,000
        & 26
        & Medium
        & {0.3472}
        & {0.1920}
        & \ccross 
        & \ccross 
        & \cmark 
        & \ccross 
        & aims to evaluate a therapy's effect, efficacy,
          safety, or outcome \\

        \BigPatent
        & Technology
        & 10,000
        & 107
        & Medium
        & {0.3757}
        & {0.2458}
         & \ccross 
        & \ccross 
        & \cmark 
        & \ccross 
        & claims to surpass or improve upon prior art \\

        \Enron
        & Business
        & 9,997
        & 324
        & Medium
        & {0.0653}
        & {0.0997}
        & \ccross
        & \ccross
        & \ccross
        & \cmark
        & was sent by a senior or high-ranking
          Enron employee \\

        \GovReport
        & Government
        & 10,000
        & 500
        & Low
        & {0.7748}
        & {0.8529}
        & \ccross 
        & \ccross 
        & \cmark 
        & \ccross
        & identifies a flaw or problem \\

        \Court
        & Legal
        & 16,902
        & 2,816
        & High
        & {0.7136}
        & {0.6782}
        & \ccross
        & \halfcheck
        & \ccross
        & \cmark
        & reverses a lower-court ruling \\

        \Legal
        & Legal
        & 489
        & 8,176
        & High
        & {0.4724}
        & {0.1881}
        & \ccross
        & \ccross
        & \ccross
        & \cmark
        & contains a covenant not to sue or
          challenge IP rights \\

        \FEVER
        & Fact verification
        & 15,981
        & 343
        & High
        & {0.8105}
        & {0.8159}
        & \cmark
        & \ccross
        & \ccross
        & \cmark
        & contains at least one document that
          clearly supports the claim \\

        \Movie
        & Film
        & 1,000
        & 23,095
        & High
        & {0.1750}
        & {0.3160}
        & \ccross
        & \cmark
        & \ccross 
        & \ccross
        & protagonist makes a critical decision based on false information \\

        \Wiki
        & Collaboration
        & 90,000
        & 54
        & Medium
        & {0.2257}
        & {0.1654}
        & \ccross
        & \cmark
        & \ccross
        & \halfcheck
        & asks another editor a genuine question
          or requests clarification \\

        \Steam
        & Review
        & 90,000
        & 80
        & Low
        & 0.4272
        & 0.4347
        & \ccross
        & \cmark
        & \ccross
        & \halfcheck
        & expresses an overall positive opinion
          of the game \\

        \bottomrule
    \end{tabular}
\end{table*}

We use OpenAI o4-mini~\cite{openai2025o4mini} for proxy task generation in \TaskCascade. All the LLMs are locally deployed by SGLang~\cite{zheng2024sglang} with chain-of-thought (CoT) reasoning disabled and a decoding temperature of $0$. Following the original paper, we use NV-Embed-v2~\cite{lee2025nvembed} as the document embedding model for \ScaleDoc.


\stitle{Experimental Configurations.}
We set the calibration document set $D'$ as 10\% of the full document set $D$ for all systems. In addition, a disjoint set of 10\% of documents is randomly drawn from $D$ as the training set for \ScaleDoc's surrogate model. 
The training hyperparameters of the lightweight surrogate model follow the original setting~\cite{DBLP:journals/corr/scaledoc}, including a 70/30 training-validation split, two training phases of 50 and 30 epochs, a learning rate of $10^{-3}$, and a batch size of 128. Threshold calibration retains the default 64 histogram bins, a five-bin smoothing window, and 33 candidate thresholds. For \TaskCascade, we follow the default three agentic proxy-generation iterations, five surrogate requests per iteration, and candidate document fraction of $\{0.1, 0.25, 0.5, 1.0\}$. 
By default, we set the accuracy target $\alpha = 0.95$ and the failure probability $\delta = 0.05$. When evaluating the Target Satisfaction Rate (TSR), we execute 100 independent runs per query on \LOTUS, \BARGAIN, and \ScaleDoc. \TaskCascade spends 1.6-40.1 hours to process a single query. We execute 1 run per query for \TaskCascade, and its TSR remains under-explored due to limited computational resources.


\stitle{Environments.} We implement the evaluation in Python. All experiments are conducted on a Linux server with two AMD EPYC 9755 128-core CPUs, 503 GB of memory, and two  NVIDIA H200 GPUs with 140GB of GPU memory.

\section{Test suite Construction}
\label{sec:benchmark-construction}

Our test suite comprises 10 dataset--predicate workloads collected
from real-world document sets, complemented by our generated 
workload variants with controlled characteristics, providing evaluation conditions beyond the distributions
represented by the original workloads.
All variants reuse existing records and predicates without
introducing additional source data.


\subsection{Datasets and Predicates}
\label{sec:datasets-and-predicates}

We evaluate the four semantic filter systems on 254,369 records
from ten datasets. Eight dataset--query workloads are adopted
from prior work~\cite{DBLP:journals/pvldb/lotus,DBLP:journals/corr/taskcascades,DBLP:journals/corr/scaledoc,DBLP:journals/pacmmod/bargain}, while we design new predicates for \Wiki and \Steam. Each dataset is paired with a natural-language binary predicate evaluated over a designated input field, requiring a \texttt{True}/\texttt{False} decision.
The datasets cover scientific and technical text
(\PubMed~\cite{dernoncourt-lee-2017-pubmed} and \BigPatent~\cite{sharma-etal-2019-bigpatent}), institutional and legal documents
(\Enron~\cite{10.1007/978-3-540-30115-8_22}, \GovReport~\cite{huang-etal-2021-efficient}, \Court~\cite{fiddler2022scotus}, and \Legal~\cite{hendrycks2021cuad}), fact verification and
narrative understanding (\FEVER~\cite{thorne-etal-2018-fever} and \Movie~\cite{movie_scripts_corpus}), and user-generated
content (\Wiki~\cite{danescu2012echoes,chang2020convokit} and \Steam~\cite{steam_reviews_2021}). Average predicate-input lengths range
from 26 words for \PubMed sentences to 23,095 words for \Movie
screenplays. The predicates cover topical, relational, evidential, and pragmatic
conditions rather than a single notion of document relevance.
Table~\ref{tab:dataset-statistics} summarizes the domains, the number of documents $|D|$, average document lengths, predicate difficulty, selectivity, and short summaries of predicates, along with dataset usage in the four original system studies. 
The symbols indicate whether the corresponding study uses the same dataset version ({\cmark}), a related dataset from the same family but with a different source or version ({\halfcheck}), or does not use the dataset ({\ccross}).


\subsection{Controlled Workload Generation}
\label{sec:selection-based-workloads}

Naturally occurring workloads reflect real-world data distributions but may not provide the variation in workload compositions needed for systematic evaluation. Therefore, we construct controlled workloads by selecting documents from collections in~\cref{sec:datasets-and-predicates}. For each oracle-model setting, we use the oracle's binary labels and predicted-label confidence to guide workload selection. The confidence of a record is defined as the token probability assigned to the oracle's final predicted label. 

\stitle{Label selectivity.}
We construct workloads with different proportions of
oracle-positive records by selecting positive and negative
records according to a specified label ratio.
This allows us to evaluate how changes in class composition
affect filtering quality, oracle reduction, and cascadeability
under a common accuracy target.

\stitle{Oracle confidence score.}
We construct four workload variants according to the oracle's
predicted-label confidence, which reflects the oracle's uncertainty in making individual binary decisions, regardless of whether the predicted label is positive or negative. The \texttt{High} and \texttt{Low} variants select the 10,000 records with the highest and lowest confidence, respectively. The \texttt{Mixed} variant combines the 5,000 highest-confidence and 5,000 lowest-confidence records. The \texttt{Uniform} variant selects 3,000 records distributed across the confidence range. These variants provide workloads with different oracle-confidence characteristics, without changing the underlying documents or predicates.

\section{Experimental Study}
\label{sec:exp}





\begin{figure}[t]
    \centering
    
    \subfigure[Brier Score ($\times 100$, $\downarrow$)] {
    \includegraphics[width=0.45\linewidth]{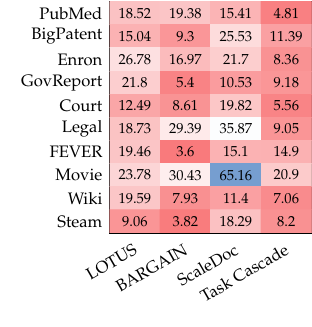}
    }
    \subfigure[ROC-AUC (\%, $\uparrow$)] {
    \includegraphics[width=0.45\linewidth]{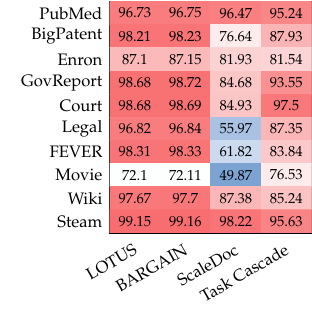}
    }
    \caption{Brier Score \& ROC-AUC of proxy scores (Qwen)}
    \label{fig:exp:AUC-BS:Qwen}
\end{figure}

\stitle{RQ \ding{182}: How well do proxy scores distinguish positive and negative documents?}


Fig.~\ref{fig:exp:AUC-BS:Qwen} reports the ROC-AUC and Brier score of the proxy. A higher AUC indicates that proxy scores more consistently rank positive records above negative records. A lower Brier score indicates a smaller mean squared difference between the scores and binary oracle labels. \LOTUS and \BARGAIN achieve nearly identical AUC and generally rank positive records above negative records more consistently than \ScaleDoc and \TaskCascade. Although \ScaleDoc achieves a high AUC on \Steam, its ranking quality is considerably lower on \Legal, \FEVER, and \Movie. \TaskCascade exhibits more workload-dependent behavior: it surpasses both small-LLM proxies on \Movie, but does not consistently retain this advantage across other workloads.

The Brier scores reveal a different aspect of proxy quality. \LOTUS and \BARGAIN exhibit substantially different Brier scores, despite their nearly identical AUC. \BARGAIN generally achieves a lower Brier score, while \LOTUS performs better on \PubMed and \Movie under both model settings. This difference partly reflects their score transformations: \LOTUS applies quantile normalization, which changes numerical score values while largely preserving their ordering, whereas \BARGAIN uses probability-derived scores. 
On workloads with low positive-label prevalence, a small Brier score can coexist with weak class separation. Brier score and AUC therefore characterize different properties of the proxy scores.


\finding{1}{}{
Small LLM proxy with the original query as input generally performs the best in ranking and distinguishing positive and negative documents, consistently outperforming the lightweight model distilled from the oracle. The task-level proxy in \TaskCascade performs well in specific workloads.
}

\stitle{RQ \ding{183}: Does proxy deliver confidence correspond to practical accuracy?}

If a proxy score $\hat{s}_i$ is a calibrated estimate of the probability that a document receives a positive oracle label, then $\mathbb{P}(y_i=1\mid\hat{s}_i=s)=s$,
where $y_i$ denotes the oracle label. Under this condition, accepting documents with $\hat{s}_i\geq\alpha$ yields
    $\mathbb{P}(y_i=1\mid\hat{s}_i\geq\alpha)
    =\mathbb{E}[\hat{s}_i\mid\hat{s}_i\geq\alpha]
    \geq\alpha.$
A corresponding relation holds for the negative acceptance region with $\hat{s}_i\leq1-\alpha$. However, the proxy scores of existing systems do not necessarily satisfy this calibration condition. Fig.~\ref{fig:exp:confidence:qwen} compares the mean proxy score in each bin with the corresponding fraction of positive documents. For the small LLM proxy, the observed positive fractions often fall below the predicted True probabilities across the score range. This suggests they tend to overestimate the predicted-label probabilities, reflecting a common problem with LLM token probabilities. In contrast, \ScaleDoc exhibits more heterogeneous relationships across workloads. On \FEVER, the observed positive fractions exceed the proxy scores across much of the intermediate score range, whereas on \Movie and \Legal, even high proxy scores correspond to relatively low positive fractions. This behavior is consistent with \ScaleDoc's scoring mechanism, which uses cosine similarities between contrastively trained document and query representations, optimized purely for class discrimination.



\finding{2}{}{
Proxy confidence is not a reliable or transferable signal of agreement with the oracle. Both small LLM and lightweight models suffer from overconfidence in their predictions. Therefore, workload- and query-specific threshold recalibration is necessary for semantic filter systems to achieve the target accuracy.
}

\newlength{\confidencefigheight}







\begin{figure}[t]
    \centering
    \setlength{\confidencefigheight}{0.37\linewidth}

    \subfigure[Small-LLM proxy]{
        \includegraphics[
            height=\confidencefigheight,
        ]{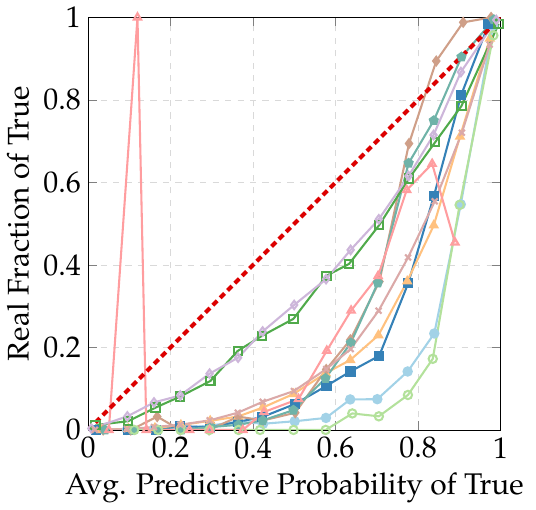}
    }%
    \hfill
    \subfigure[\ScaleDoc]{
        \includegraphics[
            height=\confidencefigheight,
        ]{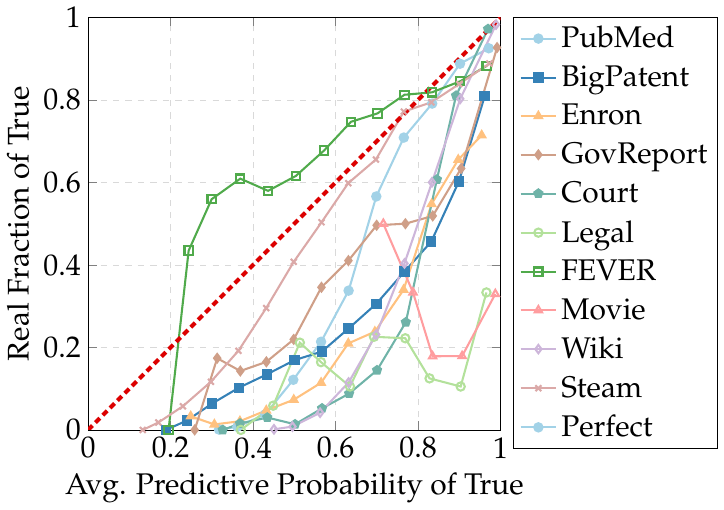}
    }

    \caption{Calibration of proxy confidence (Qwen family)}
    \label{fig:exp:confidence:qwen}
\end{figure}

\stitle{RQ \ding{184}: How much oracle-reduction opportunity does a proxy provide, and how does the two-threshold cascade realize?}

\begin{figure}
    \centering
    
    \subfigure[Qwen] {
    \includegraphics[width=0.52\linewidth]{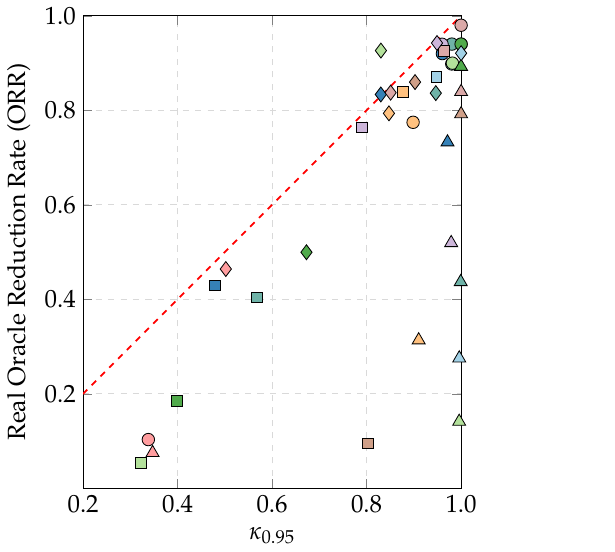}
    }
    \hspace{-8ex} \hfill
    \subfigure[Gemma] {
    \includegraphics[width=0.52\linewidth]{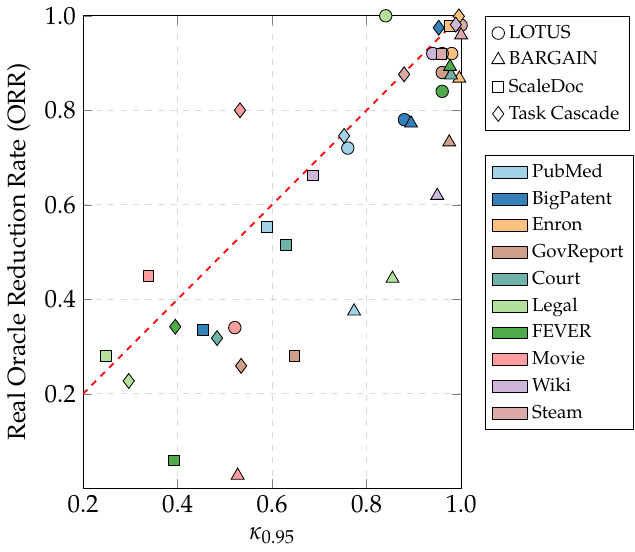}
    }
    
    \caption{Real ORR vs. $\kappa_{0.95}$ (Qwen \& Gemma families)}
    \label{fig:exp:ORR-vs-cascadabilty}
\end{figure}

\begin{figure*}
    \includegraphics[width=0.35\linewidth]{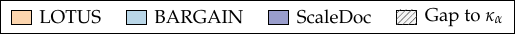}
    \par\vspace{0.5em}
    \subfigure[\PubMed] {
    \includegraphics[width=0.20\linewidth]{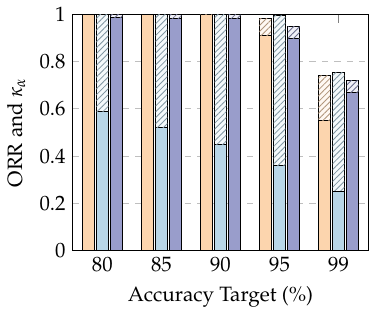}
    } 
    \hspace{-2ex}
    \subfigure[\BigPatent] {
    \includegraphics[width=0.18\linewidth]{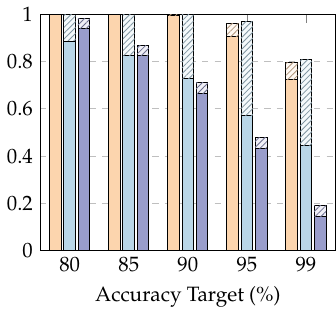}
    }
    \hspace{-2ex}
    \subfigure[\Enron] {
    \includegraphics[width=0.18\linewidth]{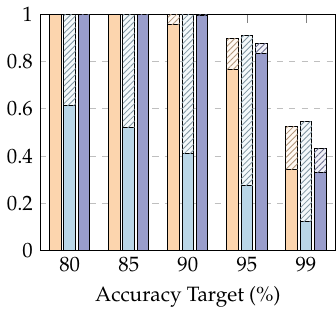}
    }
    \hspace{-2ex}
    \subfigure[\GovReport] {
    \includegraphics[width=0.18\linewidth]{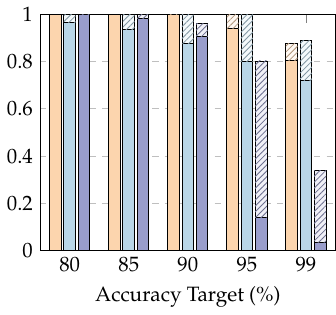}
    }
    \hspace{-2ex}
    \subfigure[\Court] {
    \includegraphics[width=0.18\linewidth]{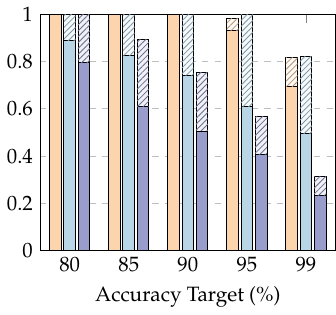}
    }
    \\
    
     \subfigure[\Legal] {
    \includegraphics[width=0.20\linewidth]{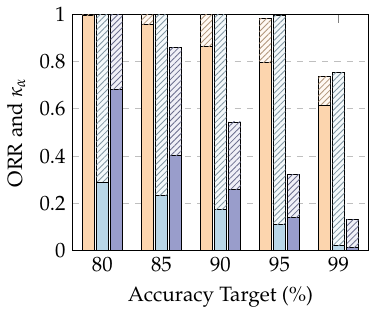}
    }
    \hspace{-2ex}
    \subfigure[\FEVER] {
    \includegraphics[width=0.18\linewidth]{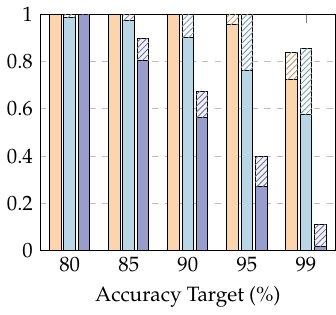}
    }
    \hspace{-2ex}
    \subfigure[\Movie] {
    \includegraphics[width=0.18\linewidth]{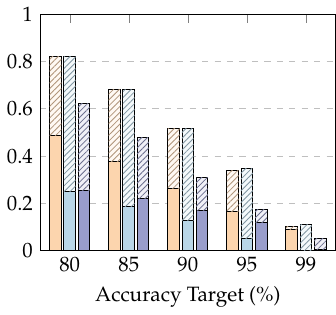}
    }
    \hspace{-2ex}
    \subfigure[\Wiki] {
    \includegraphics[width=0.18\linewidth]{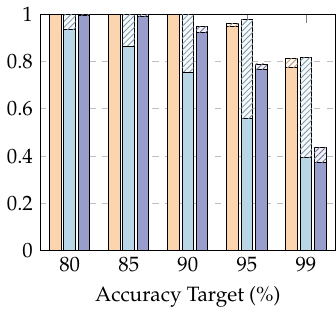}
    }
    \hspace{-2ex}
    \subfigure[\Steam] {
    \includegraphics[width=0.18\linewidth]{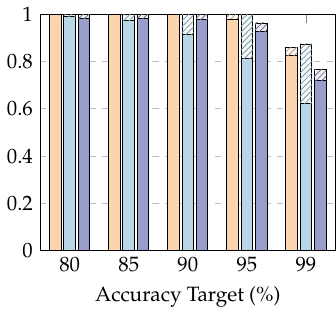}
    }
    
    \caption{Oracle reduction rate and $\kappa_\alpha$ by varying accuracy target $\alpha$ (Qwen family)}
    \label{fig:exp:ORR_kappa:Qwen}
\end{figure*}

Figs.~\ref{fig:exp:ORR-vs-cascadabilty} and~\ref{fig:exp:ORR_kappa:Qwen} compare $\kappa_{\alpha}$ with realized ORR across systems and workloads. Low ORR can arise from two distinct bottlenecks: limited
cascadeability or failure to exploit the available opportunity. At $\alpha=0.95$, seven system--workload configurations exhibit low cascadeability ($\kappa_{0.95} < 0.5)$, whereas five exhibit high cascadeability ($\kappa_{0.95} \geq 0.8$) but low realized ORR ($<0.5$).
For example, \ScaleDoc exhibits both low cascadeability
and low ORR on \Movie, indicating that the constructed proxy offers limited opportunities for
oracle reduction even with optimal thresholds. In contrast, \ScaleDoc on \GovReport and \BARGAIN on \Legal achieve low ORR despite substantially higher cascadeability. 

Systems with comparable cascadeability can also differ substantially in how much of it they realize. \LOTUS and \BARGAIN differ in $\kappa_{\alpha}$ by less than 0.02 across all ten workloads, but \LOTUS achieves at least 10 percentage points (pp) higher ORR on ten workloads, including three exceeding 40~pp. One possible contributing factor is threshold parameterization. The evaluated \BARGAIN variant uses symmetric thresholds, whereas $\kappa_\alpha$ allows the two thresholds to vary independently. This restriction may exclude asymmetric acceptance regions that would otherwise satisfy the accuracy target while achieving higher ORR. Thus, even when proxies offer comparable
oracle-reduction opportunities, differences in acceptance rules and threshold calibration can lead to different realized ORR.


\finding{3}{}{
Cascade effectiveness jointly depends on proxy quality and threshold calibration.
Low ORR can arise from two separable bottlenecks: Proxies that offer little bypassing opportunity, and calibration strategies that fail to exploit the opportunity the proxy does offer. Both bottlenecks are observed among the four evaluated systems.
}


\stitle{RQ \ding{185}: What Accuracy, ORR, and TSR does each system achieve?}

\begin{figure}
    \centering
    
    \includegraphics[width=\linewidth]{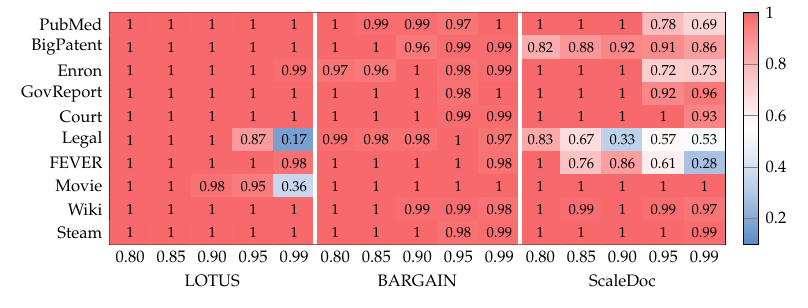}
    \caption{Target Satisfaction Rate (TSR) (Qwen family) }
    \label{fig:exp:TSR:heatmap}
\end{figure}

\begin{figure*}
    \subfigure[\PubMed] {
    \includegraphics[width=0.195\linewidth]{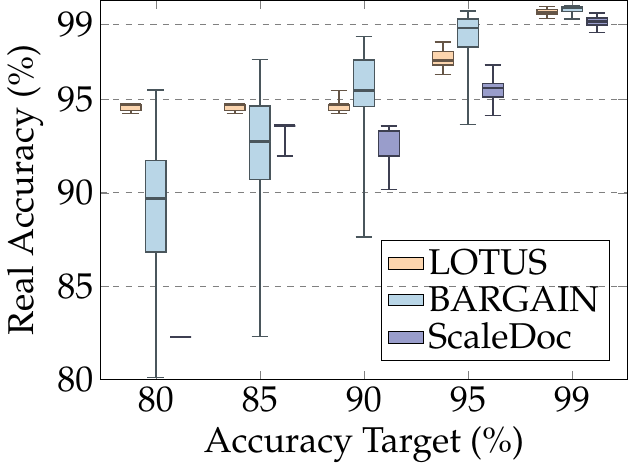}
    } 
    \hspace{-2ex}
    \subfigure[\BigPatent] {
    \includegraphics[width=0.18\linewidth]{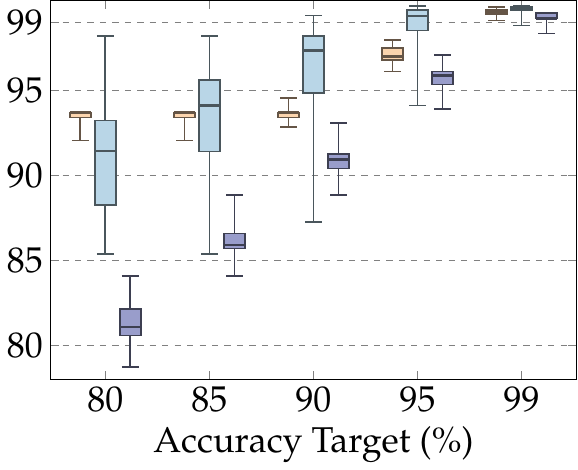}
    }
    \hspace{-2ex}
    \subfigure[\Enron] {
    \includegraphics[width=0.18\linewidth]{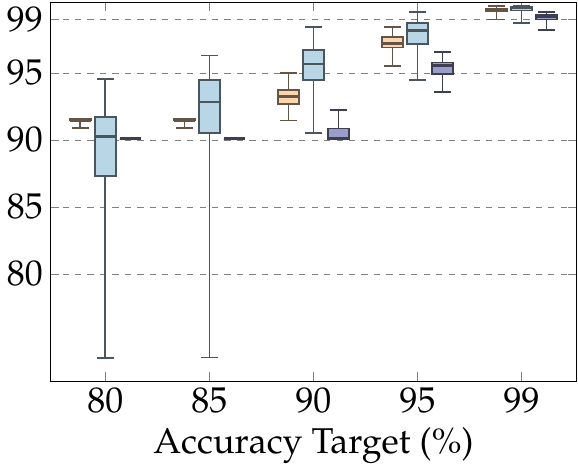}
    }
    \hspace{-2ex}
    \subfigure[\GovReport] {
    \includegraphics[width=0.18\linewidth]{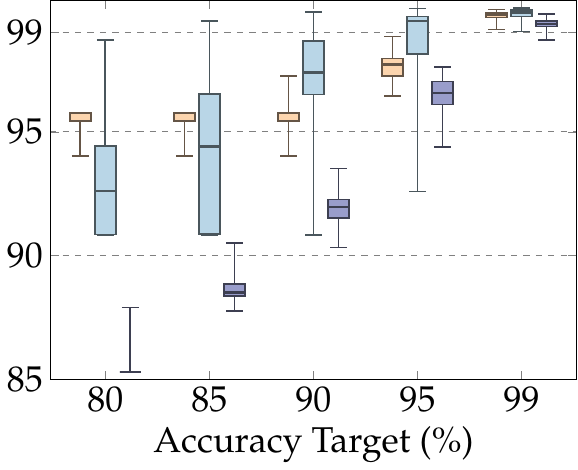}
    }
    \hspace{-2ex}
    \subfigure[\Court] {
    \includegraphics[width=0.18\linewidth]{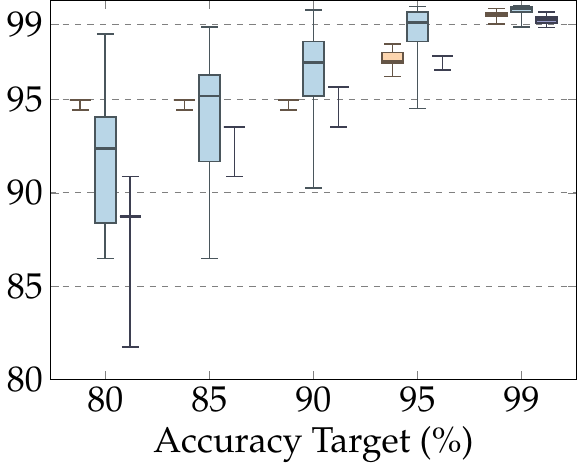}
    }
    \\
    
     \subfigure[\Legal] {
    \includegraphics[width=0.195\linewidth]{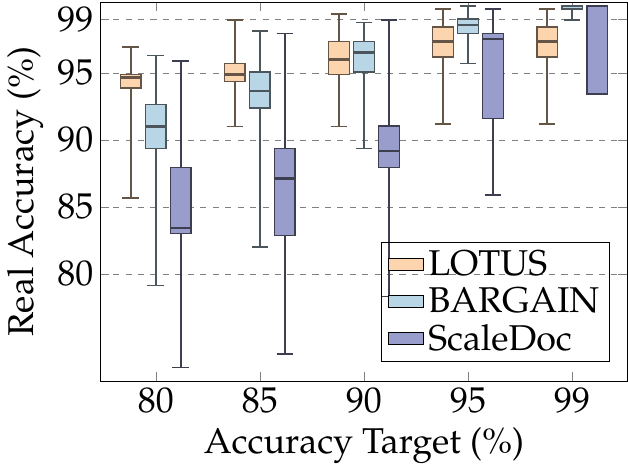}
    }
    \hspace{-2ex}
    \subfigure[\FEVER] {
    \includegraphics[width=0.18\linewidth]{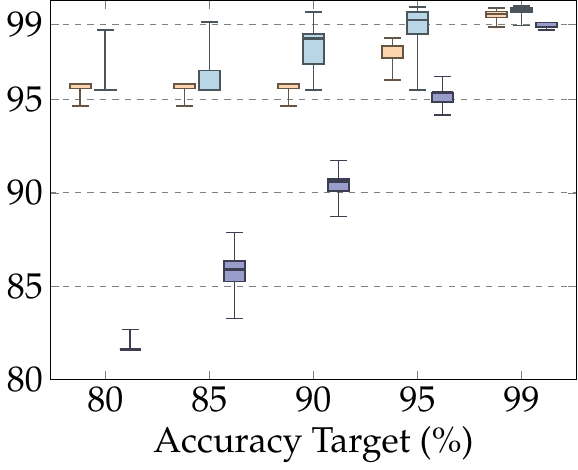}
    }
    \hspace{-2ex}
    \subfigure[\Movie] {
    \includegraphics[width=0.18\linewidth]{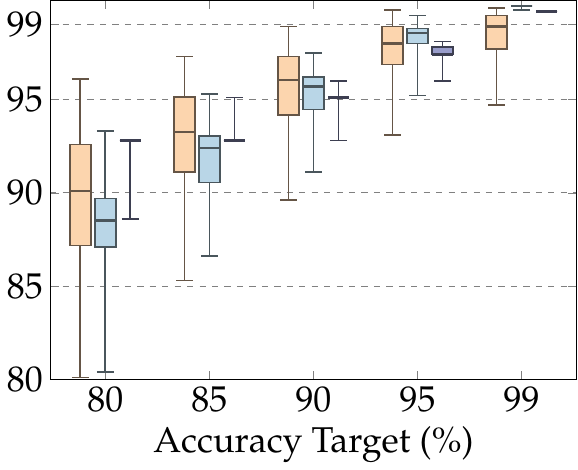}
    }
    \hspace{-2ex}
    \subfigure[\Wiki] {
    \includegraphics[width=0.18\linewidth]{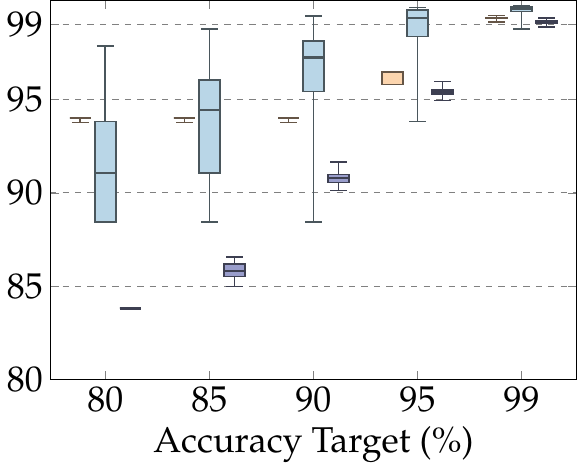}
    }
    \hspace{-2ex}
    \subfigure[\Steam] {
    \includegraphics[width=0.18\linewidth]{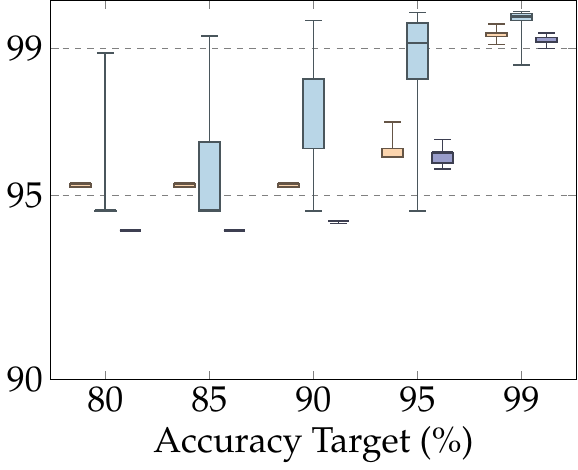}
    }

    \caption{Accuracy as the target $\alpha$ varies(Qwen family)}
    \label{fig:exp:accuracy:Qwen}
\end{figure*}

We evaluate four semantic filter systems across 10 workloads under two model families. Table~\ref{tab:qwen-overall} summarizes their performance at the default accuracy target $\alpha =0.95$. We further evaluate \LOTUS, \BARGAIN, and \ScaleDoc at accuracy targets ranging from 0.80 to 0.99. Fig.~\ref{fig:exp:TSR:heatmap} reports target satisfaction across these targets, while Fig. ~\ref{fig:exp:accuracy:Qwen} shows the distributions of achieved accuracy across repeated runs.

\LOTUS achieves high oracle reduction but is less reliable on certain challenging workloads. Across the 10 workloads, it achieves the highest mean ORR among the four systems, while maintaining a high overall TSR of 98.2\%.
However, its failures are concentrated on \Legal at the default target, and as the target increases, they become more frequent on both \Legal and \Movie. These failures are accompanied by different routing behaviors. On \Legal, \LOTUS continues to resolve a substantial fraction of records using the proxy despite frequent target violations. On \Movie, its ORR decreases considerably, yet target violations persist, showing that increasing oracle fallback does not necessarily restore target satisfaction on challenging workloads. Both workloads contain long documents and require complex semantic judgments (Table~\ref{tab:dataset-statistics}).  


\BARGAIN maintains reliable target satisfaction, but its conservative behavior often compromises oracle reduction. At the default $\alpha =0.95$, \BARGAIN achieves the highest overall TSR among systems evaluated over repeated runs. Compared with \LOTUS, its ORR is lower on every workload, yet its TSR is higher only on \Legal and \Movie. \TaskCascade also achieves higher ORR than \BARGAIN on nine workloads. This conservative behavior persists across accuracy targets. Even at relaxed targets ($\alpha=0.80$ and $0.85$), \BARGAIN frequently achieves accuracy well above the required level while maintaining considerably lower ORR than \LOTUS. As $\alpha$ increases, \BARGAIN maintains high target satisfaction at the expense of ORR, with TSR remaining at least 95\% across the evaluated configurations, consistent with the specified failure probability $\delta=0.05$.

\ScaleDoc and \TaskCascade perform well on specific workloads but exhibit inconsistent performance across datasets and model settings. At $\alpha = 0.95$, \ScaleDoc achieves a TSR of at least 95\% on only four of the ten workloads. While it combines high ORR with reliable target satisfaction on \Steam, its high ORR on \Enron is accompanied by frequent target violations. Conversely, on \Legal, it achieves
neither substantial oracle reduction nor reliable target satisfaction. This difficulty persists at stricter accuracy targets, where increased oracle fallback fails to consistently restore target satisfaction on \Legal. \TaskCascade meets the accuracy target on eight of the ten workloads. It achieves high ORR while satisfying the target on \PubMed and \Wiki, but violates the target on \BigPatent and \Legal despite substantial oracle reduction.




\finding{4}{}{ None of the evaluated systems consistently deliver both TSR and ORR across all workloads. \BARGAIN achieves the most reliable accuracy and TSR among the systems, with a compromise of ORR.
Semantic filter systems therefore favor an adaptive, workload-aware quality control design.
}

\providecommand{\SemImprove}[1]{%
  \begingroup\setlength{\fboxsep}{1pt}%
  \colorbox{cyan!15}{#1}%
  \endgroup
}

\begin{table*}[t]
\centering
\caption{%
Accuracy (\%), ORR (\%), and TSR of the four systems and proxy-switch ablations at  $\alpha=0.95$ (Qwen family). The \colorbox{Best}{best} and \colorbox{SecondBest}{second best} results among the four systems, the \colorbox{Improve}{improved} vs. both components, and the \colorbox{Fail}{unsatisfied} are highlighted.
}
\label{tab:qwen-overall}
\footnotesize
\setlength{\tabcolsep}{3.5pt}
\begin{tabular}{@{}lccc ccc cc ccc ccc ccc@{}}
\toprule
\multirow{2}{*}{\textbf{Dataset}}
& \multicolumn{3}{c}{\LOTUS}
& \multicolumn{3}{c}{\BARGAIN}
& \multicolumn{2}{c}{\TaskCascade}
& \multicolumn{3}{c}{\ScaleDoc}
& \multicolumn{3}{c}{\makecell[c]{\ScaleDoc proxy \& \\ \LOTUS calibration}}
& \multicolumn{3}{c}{\makecell[c]{\ScaleDoc proxy \& \\ \BARGAIN calibration}} \\
\cmidrule(lr){2-4}
\cmidrule(lr){5-7}
\cmidrule(lr){8-9}
\cmidrule(lr){10-12}
\cmidrule(lr){13-15}
\cmidrule(l){16-18}
& \textbf{Acc.} & \textbf{ORR} & \textbf{TSR}
& \textbf{Acc.} & \textbf{ORR} & \textbf{TSR}
& \textbf{Acc.} & \textbf{ORR}
& \textbf{Acc.} & \textbf{ORR} & \textbf{TSR}
& \textbf{Acc.} & \textbf{ORR} & \textbf{TSR}
& \textbf{Acc.} & \textbf{ORR} & \textbf{TSR} \\
\midrule
\PubMed
& 97.26 & \cellcolor{SecondBest}{90.95} & \cellcolor{Best}{100/100}
& \cellcolor{Best}{98.45} & 36.48 & \cellcolor{SecondBest}{97/100}
& \cellcolor{SecondBest}{97.42} & \cellcolor{Best}{92.08}
& 95.53 & 90.06 & \cellcolor{Fail}{78/100}
& 97.11 & 85.40 & 100/100
& 97.60 & 20.96 & \SemImprove{100/100} \\

\BigPatent
& \cellcolor{SecondBest}{97.16} & \cellcolor{Best}{90.58} & \cellcolor{Best}{100/100}
& \cellcolor{Best}{99.02} & 57.56 & \cellcolor{SecondBest}{99/100}
& \cellcolor{Fail}{94.75} & \cellcolor{SecondBest}{83.35}
& 95.65 & 43.77 & \cellcolor{Fail}{91/100}
& \cellcolor{Improve}{97.53} & 30.46 & 100/100
& 97.41 & 8.60 & 99/100 \\

\Enron
& \cellcolor{SecondBest}{97.28} & 76.59 & \cellcolor{Best}{100/100}
& \cellcolor{Best}{98.00} & 27.97 & \cellcolor{SecondBest}{98/100}
& 96.06 & \cellcolor{SecondBest}{79.35}
& 95.38 & \cellcolor{Best}{85.17} & \cellcolor{Fail}{72/100}
& \cellcolor{Improve}{97.32} & 67.11 & 100/100
& 97.01 & 7.22 & 97/100 \\

\GovReport
& \cellcolor{SecondBest}{97.67} & \cellcolor{Best}{94.06} & \cellcolor{Best}{100/100}
& \cellcolor{Best}{98.85} & 80.46 & \cellcolor{SecondBest}{98/100}
& 95.56 & \cellcolor{SecondBest}{85.97}
& 96.49 & 18.82 & \cellcolor{Fail}{92/100}
& 96.92 & 65.76 & 100/100
& 98.38 & 43.86 & 97/100 \\

\Court
& 97.15 & \cellcolor{Best}{93.21} & \cellcolor{Best}{100/100}
& \cellcolor{Best}{98.78} & 61.07 & \cellcolor{SecondBest}{99/100}
& 96.11 & \cellcolor{SecondBest}{83.64}
& \cellcolor{SecondBest}{97.22} & 40.66 & \cellcolor{Best}{100/100}
& 96.76 & 46.03 & 100/100
& 98.22 & 30.81 & 100/100 \\

\Legal
& \cellcolor{SecondBest}{97.17} & \cellcolor{SecondBest}{81.54} & \cellcolor{Fail}{87/100}
& \cellcolor{Best}{98.39} & 11.19 & \cellcolor{Best}{100/100}
& \cellcolor{Fail}{92.84} & \cellcolor{Best}{92.64}
& 95.11 & 18.51 & \cellcolor{Fail}{57/100}
& 95.94 & 26.11 & \cellcolor{Fail}{74/100}
& \SemImprove{99.98} & 0.02 & 100/100 \\

\FEVER
& \cellcolor{SecondBest}{97.32} & \cellcolor{Best}{95.53} & \cellcolor{Best}{100/100}
& \cellcolor{Best}{98.90} & \cellcolor{SecondBest}{76.16} & \cellcolor{Best}{100/100}
& 97.04 & 49.95
& 95.19 & 28.54 & \cellcolor{Fail}{61/100}
& 96.60 & 27.46 & 97/100
& 97.75 & 20.05 & 100/100 \\

\Movie
& \cellcolor{SecondBest}{97.85} & \cellcolor{SecondBest}{17.23} & \cellcolor{SecondBest}{95/100}
& \cellcolor{Best}{98.29} & 5.06 & \cellcolor{Best}{100/100}
& 95.30 & \cellcolor{Best}{46.40}
& 97.39 & 11.94 & \cellcolor{Best}{100/100}
& 97.19 & 9.68 & \cellcolor{Fail}{85/100}
& \cellcolor{Improve}{99.42} & 0.74 & 100/100 \\

\Wiki
& \cellcolor{SecondBest}{96.16} & \cellcolor{Best}{94.87} & \cellcolor{Best}{100/100}
& \cellcolor{Best}{98.91} & 56.17 & \cellcolor{SecondBest}{99/100}
& 95.00 & \cellcolor{SecondBest}{94.24}
& 95.45 & 76.70 & \cellcolor{SecondBest}{99/100}
& 96.08 & 72.86 & 100/100
& 98.36 & 44.71 & 97/100 \\

\Steam
& \cellcolor{SecondBest}{96.11} & \cellcolor{Best}{97.99} & \cellcolor{Best}{100/100}
& \cellcolor{Best}{98.80} & 81.67 & \cellcolor{SecondBest}{98/100}
& 95.25 & 83.73
& 96.06 & \cellcolor{SecondBest}{92.80} & \cellcolor{Best}{100/100}
& \cellcolor{Improve}{96.17} & 92.86 & 100/100
& 98.51 & 37.86 & 99/100 \\

\midrule
\textbf{Overall}
& \cellcolor{SecondBest}{97.11} & \cellcolor{Best}{83.26} & \cellcolor{SecondBest}{982/1000}
& \cellcolor{Best}{98.64} & 49.38 & \cellcolor{Best}{988/1000}
& 95.53 & \cellcolor{SecondBest}{79.13}
& 95.95 & 50.70 & \cellcolor{Fail}{850/1000}
& 96.76 & 52.37 & 956/1000
& 98.26 & 21.48 & \cellcolor{Improve}{989/1000} \\
\bottomrule
\end{tabular}%
\end{table*}

\stitle{RQ \ding{186}:  What is the token cost and end-to-end execution latency?}

\begin{figure*}[t]
    \centering
    \includegraphics[width=0.55\linewidth]{%
        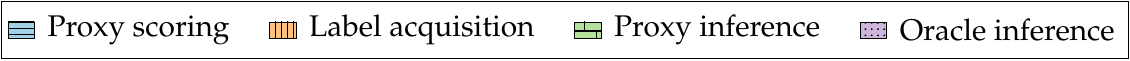}
    \par\vspace{0.35em}
    \subfigure[\PubMed]{%
        \includegraphics[width=0.25\linewidth]{%
            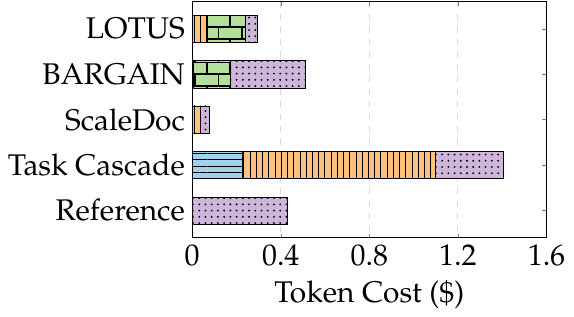}%
    }%
    \hfill
    \subfigure[\BigPatent]{%
        \includegraphics[width=0.17\linewidth]{%
            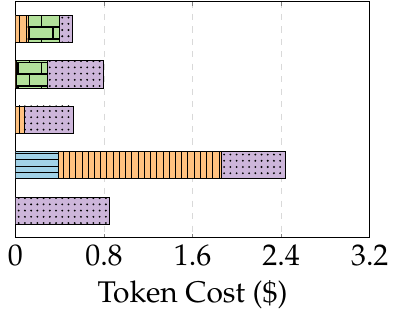}%
    }%
    \hfill
    \subfigure[\Enron]{%
        \includegraphics[width=0.17\linewidth]{%
            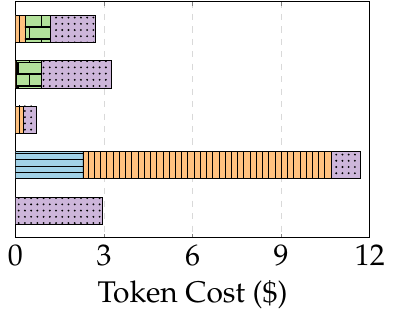}%
    }%
    \hfill
    \subfigure[\GovReport]{%
        \includegraphics[width=0.17\linewidth]{%
            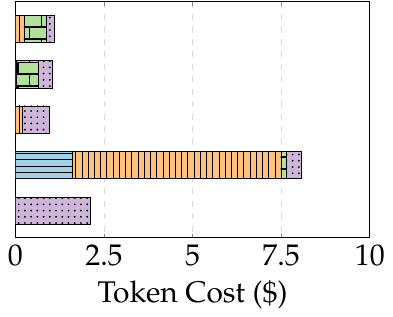}%
    }%
    \hfill
    \subfigure[\Court]{%
        \includegraphics[width=0.17\linewidth]{%
            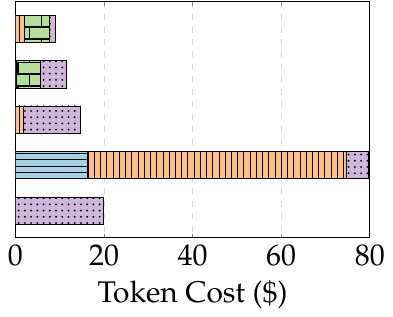}%
    }%
    \par\vspace{0.3em}
    \subfigure[\Legal]{%
        \includegraphics[width=0.25\linewidth]{%
            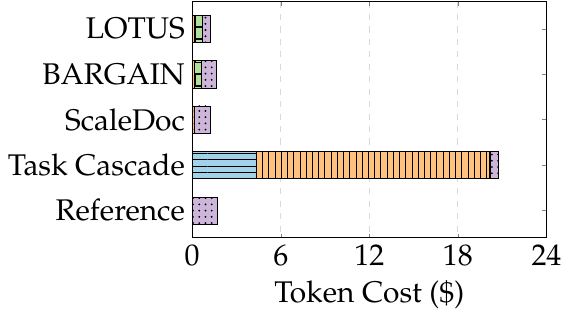}%
    }%
    \hfill
    \subfigure[\FEVER]{%
        \includegraphics[width=0.17\linewidth]{%
            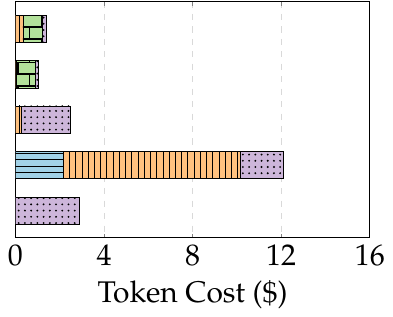}%
    }%
    \hfill
    \subfigure[\Movie]{%
        \includegraphics[width=0.17\linewidth]{%
            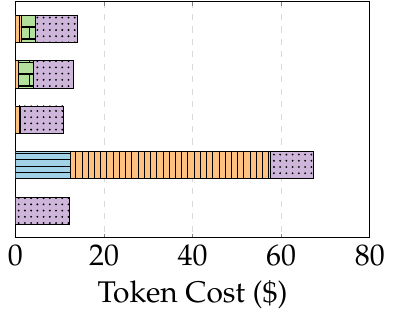}%
    }%
    \hfill
    \subfigure[\Wiki]{%
        \includegraphics[width=0.17\linewidth]{%
            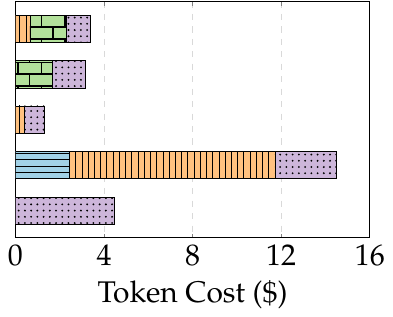}%
    }%
    \hfill
    \subfigure[\Steam]{%
        \includegraphics[width=0.17\linewidth]{%
            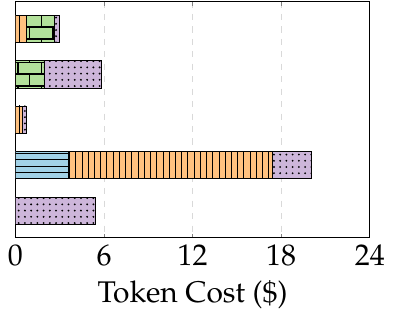}%
    }%
    
    \par\vspace{0.2em}
    \caption{Token consumption (\$) (Qwen family)}
    \label{fig:exp:token_cost:Qwen}
\end{figure*}

\begin{figure*}
    \centering
    \includegraphics[width=0.35\linewidth]{%
        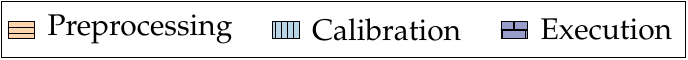}
    \par\vspace{0.35em}
    \subfigure[\PubMed]{%
        \includegraphics[width=0.25\linewidth]{%
            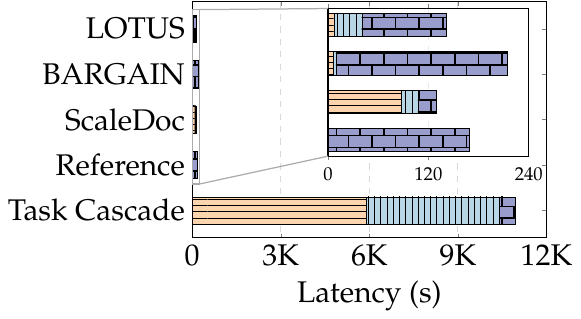}%
    }%
    \hfill
    \subfigure[\BigPatent]{%
        \includegraphics[width=0.17\linewidth]{%
            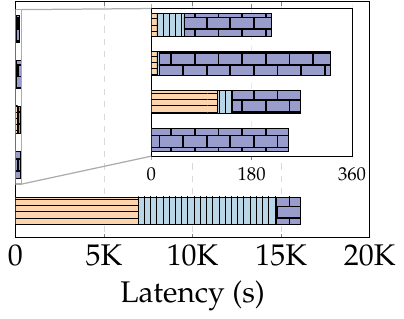}%
    }%
    \hfill
    \subfigure[\Enron]{%
        \includegraphics[width=0.17\linewidth]{%
            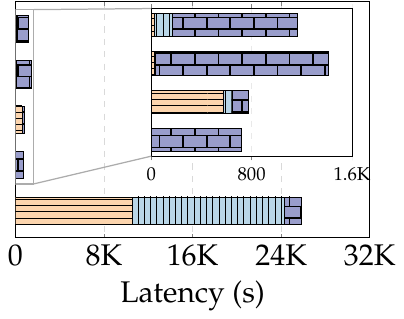}%
    }%
    \hfill
    \subfigure[\GovReport]{%
        \includegraphics[width=0.17\linewidth]{%
            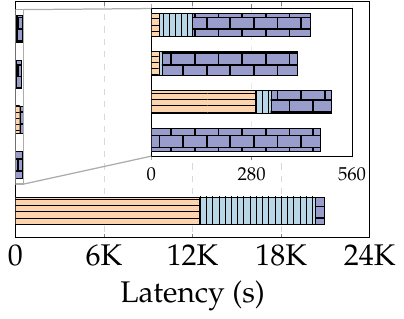}%
    }%
    \hfill
    \subfigure[\Court]{%
        \includegraphics[width=0.17\linewidth]{%
            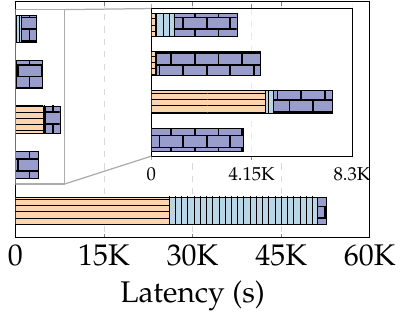}%
    }%
    \par\vspace{0.3em}
    \subfigure[\Legal]{%
        \includegraphics[width=0.25\linewidth]{%
            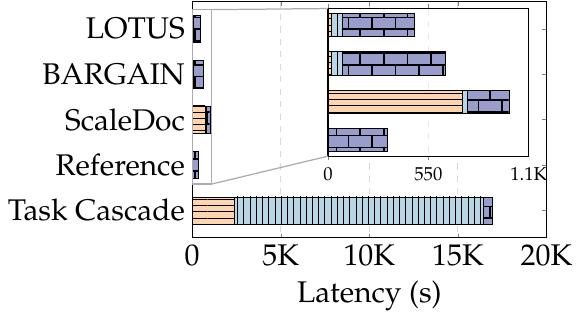}%
    }%
    \hfill
    \subfigure[\FEVER]{%
        \includegraphics[width=0.17\linewidth]{%
            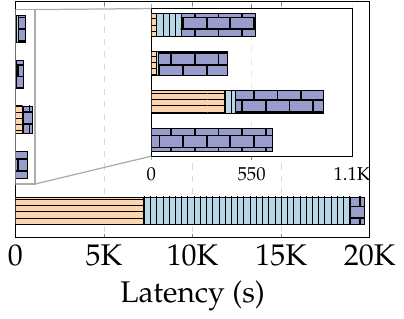}%
    }%
    \hfill
    \subfigure[\Movie]{%
        \includegraphics[width=0.17\linewidth]{%
            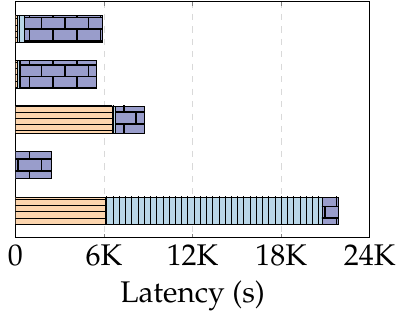}%
    }%
    \hfill
    \subfigure[\Wiki]{%
        \includegraphics[width=0.17\linewidth]{%
            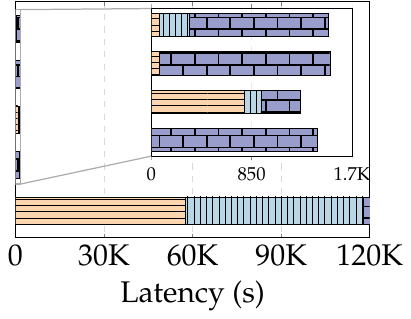}%
    }%
    \hfill
    \subfigure[\Steam]{%
        \includegraphics[width=0.17\linewidth]{%
            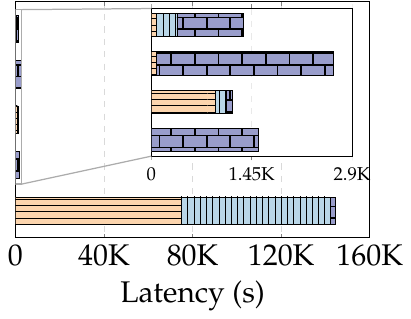}%
    }%
    
    \par\vspace{0.2em}
    \caption{End-to-end latency (seconds)  (Qwen family)}
    \label{fig:exp:latency:Qwen}
\end{figure*}

Fig.~\ref{fig:exp:token_cost:Qwen} compares the token costs, while Fig.~\ref{fig:exp:latency:Qwen} reports the end-to-end latency. The reference invokes the oracle directly on every document. 
Cost estimates only include model invocations for calibration and execution.
\LOTUS and \BARGAIN show that higher oracle reduction does not necessarily translate into proportional cost or latency savings. Although \LOTUS achieves higher ORR across all 10 workloads, it incurs higher token cost on four and longer latency on three. On \GovReport and \FEVER, \LOTUS is both more expensive and slower than \BARGAIN, indicating that proxy processing and calibration overhead can offset the savings from reduced oracle invocations. \ScaleDoc achieves lower token cost than oracle-only execution on all ten workloads, yet reduces end-to-end latency on only three. On \Court and \FEVER, frequent oracle fallback diminishes its cost advantage. Moreover, preprocessing accounts for a substantial share of its end-to-end latency, particularly on \Legal, offsetting the benefits of inexpensive proxy inference. \TaskCascade incurs substantial cascade-construction overhead that outweighs its inference savings.
Despite reducing oracle calls during execution, its reported cost and latency exceed those of oracle-only execution across all 10 workloads, with token costs 2.9-12.3$\times$ and end-to-end latencies 8.9-93.1$\times$ those of the reference. Candidate task evaluation contributes substantially to its token cost, while preprocessing and calibration dominate its latency.


\finding{5}{}{ ORR alone is not a comprehensive measure of efficiency. Proxy inference, calibration labeling, and cascade construction can offset or even outweigh the savings from oracle reduction. The lightweight surrogate of \ScaleDoc is often token-efficient despite only moderate oracle reduction, but preprocessing can dominate its end-to-end latency on certain workloads. The cost of agentic multi-stage proxy construction even surpasses the cost of oracle invocations. 
}

\begin{figure}
    \centering
    
    \subfigure[F1 of \Wiki] {
    \includegraphics[width=0.46\linewidth]{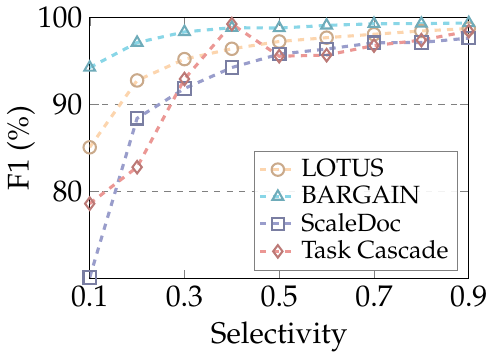}
    }
    \hspace{-3ex} \hfill
    \subfigure[F1 of \Steam] {
    \includegraphics[width=0.46\linewidth]{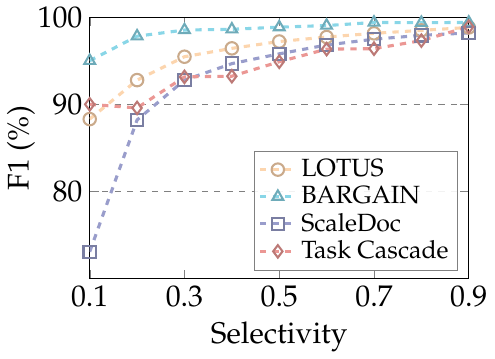}
    }
    \\
    \subfigure[Accuracy of \Wiki] {
    \includegraphics[width=0.46\linewidth]{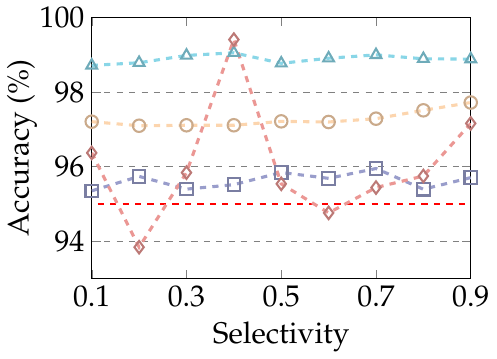}
    }
    \hspace{-3ex} \hfill
    \subfigure[Accuracy of \Steam] {
    \includegraphics[width=0.46\linewidth]{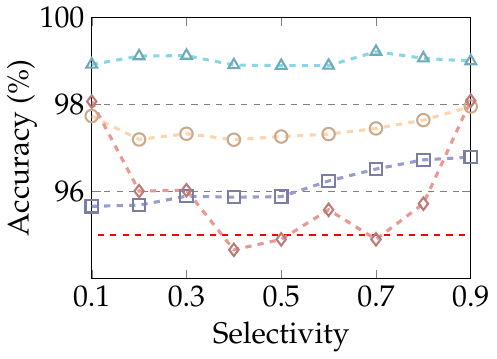}
    }
    \\
     \subfigure[ORR and $\kappa_{0.95}$ of \Wiki] {
    \includegraphics[width=0.46\linewidth]{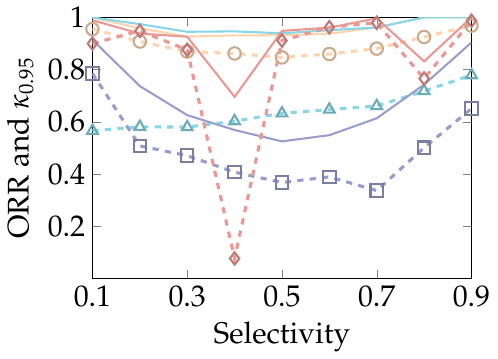}
    }
    \hspace{-3ex} \hfill
    \subfigure[ORR and $\kappa_{0.95}$ of \Steam] {
    \includegraphics[width=0.46\linewidth]{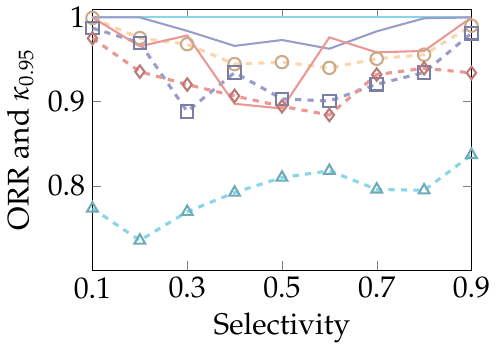}
    }
    
    \caption{Sensitivity to query selectivities (Qwen family)}
    \label{fig:exp:workload_selectivity:Qwen}
\end{figure}

\stitle{RQ \ding{187}:  How does performance vary with query selectivity?}

Fig.~\ref{fig:exp:workload_selectivity:Qwen} shows that F1 generally increases with selectivity, with the largest differences between systems occurring when positive records are rare. At a selectivity of 0.1, the F1 gap across systems exceeds 20~pp on both \Wiki and \Steam, but narrows to less than 3~pp at a selectivity of 0.9. \BARGAIN maintains relatively high F1 at low selectivity, whereas \LOTUS and \ScaleDoc exhibit more pronounced declines. Despite these differences, the achieved accuracy of \LOTUS and \BARGAIN remains relatively stable across selectivities. This discrepancy partly reflects the sensitivity of F1 to positive-class prevalence, as minority-class errors can substantially reduce F1 while contributing relatively little to overall accuracy loss. Moreover, \BARGAIN consistently achieves lower ORR than \LOTUS, suggesting that its higher F1 may partly result from more frequent oracle fallback rather than stronger proxy discrimination alone. 

Selectivity also affects the oracle-reduction opportunities available under a fixed accuracy constraint. On \Wiki, \ScaleDoc exhibits substantially lower $\kappa_{0.95}$ near balanced selectivity, dropping to approximately 0.55, with realized ORR following a similar pattern. In contrast, its cascadeability on \Steam remains approximately 0.95 across selectivities under both model settings. Part of this variation arises from the class composition of the workload: selectivity determines both how much each class contributes to overall ORR and how heavily its errors count toward the accuracy budget. When one class dominates, resolving a large fraction of that class can provide substantial oracle reduction, while errors in the minority class contribute less to overall accuracy loss. At more balanced ratios, neither class contributes negligibly to the overall error in determining feasible acceptance regions.

\finding{6}{}{ Selectivity changes which errors an accuracy target tolerates and how much opportunity a proxy can provide. Under class imbalance, minority-class errors weigh less in overall accuracy, so that high cascadeability or ORR may reflect uneven use of the accuracy budget instead of the strong discrimination of the system.
}

\stitle{RQ \ding{188}:  How does performance vary across different oracle confidence score distributions?}

\begin{figure}[!t]
    \centering
    
    \includegraphics[width=0.88\linewidth]{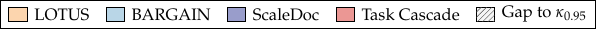}
    \hspace{-3ex}
    \subfigure[F1 of \Wiki] {
    \includegraphics[width=0.46\linewidth]{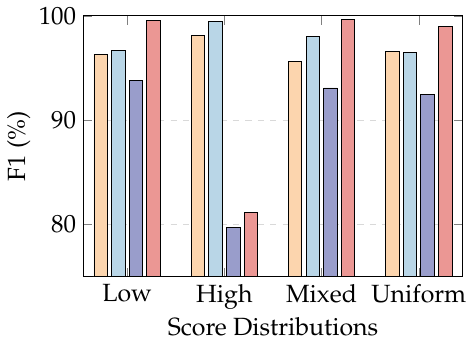}
    }
    \hspace{-3ex} \hfill
    \subfigure[F1 of \Steam] {
    \includegraphics[width=0.46\linewidth]{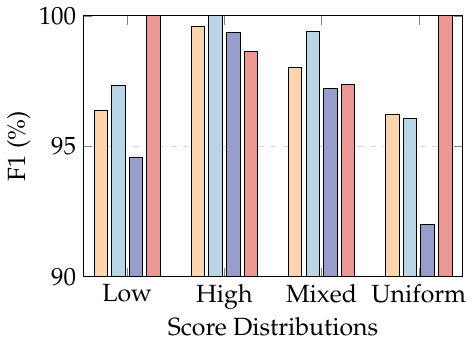}
    }
    \\
    \subfigure[Accuracy of \Wiki] {
    \includegraphics[width=0.46\linewidth]{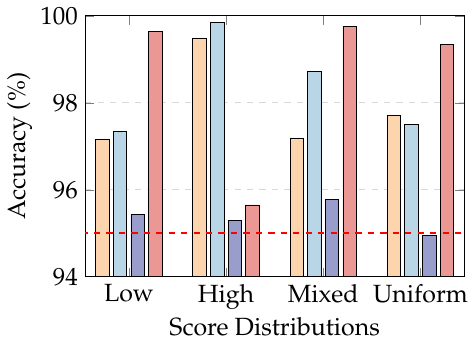}
    }
    \hspace{-3ex} \hfill
    \subfigure[Accuracy of \Steam] {
    \includegraphics[width=0.46\linewidth]{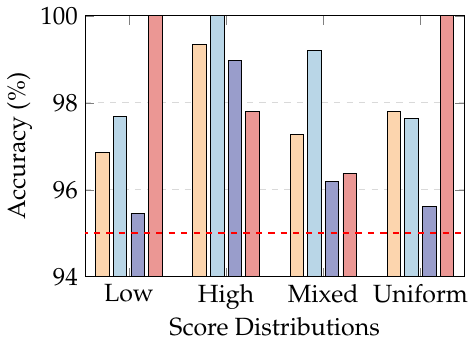}
    }
    \\
     \subfigure[ORR and $\kappa_{0.95}$ of \Wiki] {
    \includegraphics[width=0.46\linewidth]{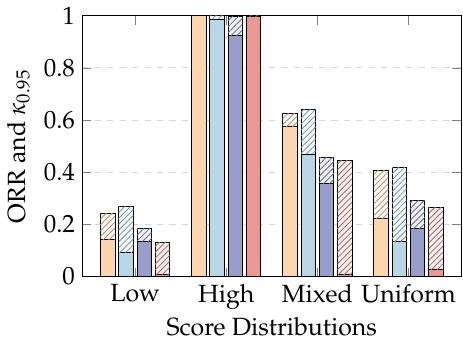}
    }
    \hspace{-3ex} \hfill
    \subfigure[ORR and $\kappa_{0.95}$ of \Steam] {
    \includegraphics[width=0.46\linewidth]{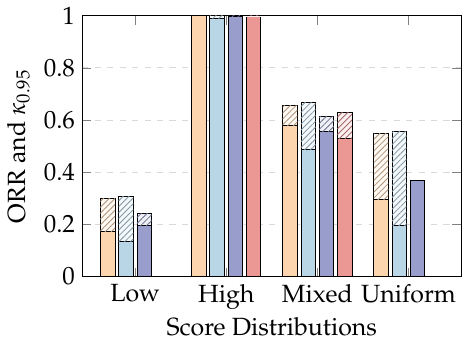}
    }
    
    \caption{Sensitivity by varying oracle confidence score distributions (Qwen family)}
    \label{fig:exp:workload_score:Qwen}
\end{figure}

Fig.~\ref{fig:exp:workload_score:Qwen} compares system performance across the confidence variants defined in \cref{sec:selection-based-workloads}. The \texttt{High} workloads exhibit near-maximal $\kappa_{0.95}$ across all four systems, whereas \texttt{Low} workloads offer substantially less opportunity for oracle reduction. Across \Wiki and \Steam, all eight system--workload configurations under \texttt{High} achieve $\kappa_{0.95}>0.95$, while all eight under \texttt{Low} remain below 0.35. This pattern also holds for \ScaleDoc and \TaskCascade, whose proxy designs do not rely on the oracle's confidence scores. In contrast, \texttt{Mixed} and \texttt{Uniform} workloads exhibit more varied cascadeability across systems and workloads. As a potentially useful signal, oracle uncertainty has yet to be explicitly exploited in current system design. Notably, high cascadeability does not necessarily imply high F1: on \texttt{High} \Wiki, \ScaleDoc and \TaskCascade achieve F1 scores around 80\%, compared with over 98\% for \LOTUS and \BARGAIN, despite near-maximal $\kappa_{0.95}$.



Beyond the available opportunities, systems also differ in how effectively they realize cascadeability. On \texttt{High} workloads, realized ORR generally approaches $\kappa_{0.95}$, whereas larger realization gaps emerge under other confidence distributions. \LOTUS generally exhibits smaller realization gaps than \BARGAIN. In particular, \BARGAIN's realization gap on \texttt{Uniform} exceeds that on \texttt{Mixed} by more than 10~pp on both \Wiki and \Steam. In contrast, \ScaleDoc nearly reaches its cascadeability on \texttt{Uniform}, while \TaskCascade leaves substantial opportunities unrealized on \texttt{Mixed} \Wiki, where its ORR remains near zero despite nontrivial cascadeability. These differences may partly arise from how systems sample records and estimate proxy quality. \BARGAIN orders the documents according to the proxy confidence and uses adaptive sampling to certify their quality. Compared with \texttt{Mixed} workloads, \texttt{Uniform} workloads include records beyond the confidence extremes,
potentially introducing more heterogeneous error patterns that make expanded acceptance regions difficult to certify. \ScaleDoc instead uses stratified sampling across score bins and reconstructs the score distributions through
interpolation and smoothing, potentially contributing to its
smaller realization gap on certain \texttt{Uniform}
workloads.

\finding{7}{}{ The underlying distribution of LLM confidence strongly influences the cascade opportunity for every proxy. Workloads with high LLM confidence are nearly fully cascadeable, whereas low LLM confidence leaves little cascade opportunity for existing proxies. Away from the high-confidence regime, realization gaps also widen in system-specific ways. 
}

\stitle{RQ \ding{189}:  What if we replace surrogate models or calibration procedures across systems?}

We combine \ScaleDoc's proxy with the calibration and threshold selection procedures of \LOTUS and \BARGAIN to examine how these components perform across frameworks. Table~\ref{tab:qwen-overall} reports the results at $\alpha = 0.95$. Despite using the same underlying proxy scores, the three calibration procedures yield substantially different trade-offs between target satisfaction and oracle reduction. Compared with native \ScaleDoc, \LOTUS calibration substantially improves overall TSR by 10.6~pp while maintaining comparable ORR, whereas \BARGAIN calibration improves TSR by 13.9~pp at the expense of a 29.2~pp reduction in ORR. These differences are also workload-dependent: \LOTUS calibration improves both TSR and ORR on \GovReport and \Legal, but reduces both on \Movie. In contrast, both \LOTUS and \BARGAIN achieve lower ORR with \ScaleDoc's proxy than with their native proxies, across all 10 workloads, without materially improving overall target satisfaction. These results indicate that calibration procedures can substantially improve the reliability of a fixed proxy, but their effectiveness does not necessarily transfer across different proxy designs.



\finding{8}{}{ Proxy construction and threshold calibration are not interchangeable modules, which should be designed and evaluated jointly.
}

\section{Related Work}
 \label{sec:relatedwork}

\stitle{LLM-powered Semantic Filtering.}
Semantic filtering is fundamental to LLM-powered data processing. Palimpzest~\cite{liu2025palimpzest}, DocETL~\cite{shankar2025docetl}, and SEMA~\cite{DBLP:journals/corr/abs-2603-11622}
support natural-language predicates in declarative pipelines with optimizations such as query rewriting and operator reordering. Existing approaches reduce filtering costs either by lowering per-record inference costs or by leveraging LLM judgments across records and queries.  FrugalGPT~\cite{chen2024frugalgpt} employs model cascades for selective LLM invocation. Chung et al.~\cite{chung2026proxy} train classifiers to replace selected LLM-based semantic operators, while JEVDB~\cite{wang2026jevdb} employs Jev~\cite{typesafe2026jev}, a non-generative decision model that avoids autoregressive decoding. Other approaches exploit sampled or previously
obtained LLM judgments. UQE~\cite{dai2024uqe} iteratively trains a classifier on LLM-labeled records to identify likely positive records for further LLM evaluation under a fixed inference budget. CSV~\cite{DBLP:journals/corr/abs-2603-04799} uses clustering, sampling, and voting to infer labels for semantically similar records, recursively refining ambiguous clusters. Kim et al.~\cite{kim2026fastfilter} extend CSV with
an adaptive two-phase strategy that switches to proxy-based filtering when clustering is insufficient. SemViews~\cite{hassanzadeh2026semviews} reuses labels from previous queries and obtains fresh oracle samples to certify result quality.

\stitle{Approximate Query Processing.} Approximate query processing (AQP) studies the trade-offs among efficiency, result accuracy, and statistical guarantees~\cite{chaudhuri2017aqp}. AQP systems~\cite{hellerstein1997online, acharya1999aqua, haas1999ripple, agarwal2013blinkdb, li2016wanderjoin, kandula2016quickr, park2018verdictdb, peng2018aqpp} accelerate relational analytics through sampling and precomputed summaries, providing statistical error estimates. AQP also extends to expensive ML inference.  Existing systems accelerate video filtering with inexpensive models~\cite{DBLP:journals/pvldb/KangEABZ17, DBLP:conf/sigmod/LuCKC18}
and support approximate aggregation using sampled oracle evaluations~\cite{DBLP:journals/pvldb/KangBZ19,DBLP:journals/pvldb/KangGBHSZ21,DBLP:journals/pvldb/RussoHK0Z23}. For approximate selection, SUPG~\cite{DBLP:journals/pvldb/KangGBHZ20} provides precision or recall guarantees under a fixed oracle budget, while Ding et al.~\cite{DBLP:journals/pvldb/DingAL22} use proxy-based distance rankings with probabilistic error models or oracle sampling to meet such targets.

\stitle{Benchmarks for Semantic Query Processing.}
SemBench~\cite{lao2026sembench} evaluates semantic and relational operators across text, image, and audio using existing annotated datasets. Bench-U~\cite{deng2026benchu} constructs relational tables by manually extracting structured attributes from unstructured documents across six domains. It compares LLM-based analytical query results over the documents with those from equivalent relational queries. LROBench~\cite{su2026lrobench} constructs SQL-like queries with LLM-enhanced operators over relational datasets, including BIRD~\cite{li2023bird}, Magellan~\cite{dasMagellanRepository}, NextiaJD~\cite{flores2021nextiajd}, and SANTOS~\cite{khatiwada2023santos}. Expert-authored queries with human-verified answers cover individual operators and their combinations. Same Query, Different Answer~\cite{hassanzadeh2026variance} repeatedly evaluates workloads from SemBench, Bench-U, and LROBench, showing that LLM output variability at temperature zero can affect benchmark reliability. Existing benchmarks primarily evaluate end-to-end semantic query processing across diverse operators and real-world workloads. In contrast, our benchmark examines semantic filters, isolating the components for analysis and using controlled workloads to provide a fine-grained understanding of system performance and its bottlenecks.

\section{Future Research Directions}

The benchmark covers single-predicate semantic filters over static document collections. We identify the five orthogonal directions beyond this scope.

\stitle{Long documents exceed the context window.} Documents exceeding the context window require chunked oracle evaluation and possibly aggregation to obtain reference labels. \TaskCascade has explored chunk-aware cascades, which exploit predicate structure and prefix sharing across reordered chunks. 
Open problems include effective segmentation and aggregation, certifying document-level accuracy from correlated chunk scores, and supporting asymmetric acceptance regions.
The evaluation needs workloads that vary in length and evidence placement.

\stitle{Reasoning-intensive predicates.} Complicated predicates motivate oracles or proxies that use chain-of-thought (CoT) reasoning~\cite{DBLP:conf/nips/Wei0SBIXCLZ22}. Introducing CoT poses two challenges for semantic filter processing. 
First, the token cost of CoT is charged by the relatively expensive output tokens. Systems favor a higher ORR with appropriate CoT compression techniques~\cite{DBLP:journals/corr/abs-2602-08324, DBLP:conf/acl/LiLZLMLWS26, DBLP:conf/iclr/LiZZWXCZX26} for efficient processing.
Second, the reasoning trace of small LLM surrogates can degenerate substantially, frequently becoming trapped in overthinking and self-reflection loops; hence, systems may require a mechanism for transferring the CoT from the oracle to small LLM surrogates. 

\stitle{Multimodal semantic filters.} Images, audio, video, and scanned pages change both the cost structure and the choice of proxy.
Lightweight per-query proxies have been proposed to accelerate video analytics~\cite{DBLP:journals/pvldb/KangEABZ17, DBLP:conf/sigmod/LuCKC18}, and their design is similar to \ScaleDoc. Whether a multimodal LLM proxy also outperforms these lightweight proxies remains an open question. 

\stitle{Evolving collections and workload drift.} In production-level applications, document collections evolve, and query predicates recur, while the evaluated systems calibrate thresholds once for a fixed collection and predicate.
A guarantee certified on $D$ does not necessarily hold for $D \cup \Delta D$, because data drift may invalidate calibrated thresholds.
Since proxy scores are computed for every record, score shifts are observable without oracle invocations and could trigger targeted recalibration. Systems entail a distribution shift detection and incremental recalibration mechanism to dynamically maintain the cascade framework with oracle label reuse. 

\stitle{Semantic joins.} \LOTUS supports semantic joins, which apply a predicate to document pairs. As the pair space is a Cartesian product, processing every pair with a surrogate model becomes prohibitively inefficient. Therefore, cascades need a candidate-generation stage, such as embedding-based blocking. This candidate-generation stage introduces challenges for a principled guarantee of quality targets.
True positives may be lost during candidate-generation, which is invisible to threshold calibration unless the sampling covers the full pair space. However, true positives are typically sparse in the full pair space.

\section{Conclusion}
\label{sec:conclusion}
LLM-powered semantic filters are expensive when an oracle LLM evaluates every document individually. Proxy cascades reduce this cost only if an inexpensive surrogate model can resolve documents without violating a quality target. 
We present a unified benchmark that places four representative systems in a common proxy cascade framework and evaluates them under a common protocol. The test suite covers 10 real workloads across two LLM families, along with derivative synthetic workloads with controlled query selectivities and oracle confidence score distributions. We investigate 8 research questions, spanning proxy quality, end-to-end performance, robustness, and component ablations, and introduce 5 orthogonal future research directions.

\balance
\bibliographystyle{ACM-Reference-Format}
\bibliography{ref}



\clearpage
\nobalance
\appendix

\setcounter{section}{0}
\renewcommand{\thesection}{\Alph{section}}
\renewcommand{\thesubsection}{\thesection.\arabic{subsection}}

\section{Additional Experimental Results}

\begin{figure}[htp]
    \centering
    
    \subfigure[Brier Score ($\downarrow$)] {
    \includegraphics[width=0.45\linewidth]{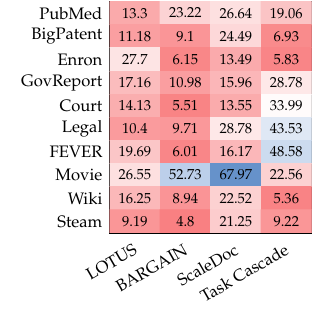}
    }
    \subfigure[AUC-ROC ($\uparrow$)] {
    \includegraphics[width=0.45\linewidth]{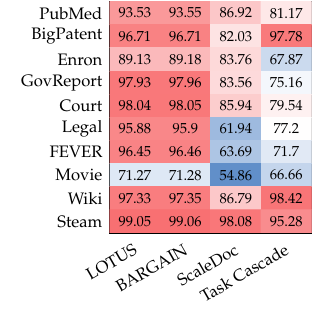}
    }
    \caption{Brier Score \& AUC-ROC of proxy scores (Gemma)}
    \label{fig:exp:AUC-BS:Gemma}
\end{figure}

\label{app:experimental-details}

\begin{figure}[htp]
    \centering
    \setlength{\confidencefigheight}{0.37\linewidth}

    \subfigure[Small-LLM proxy]{
        \includegraphics[
            height=\confidencefigheight,
        ]{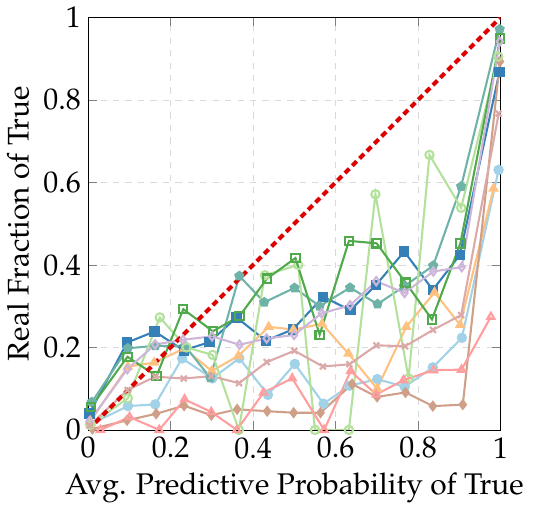}
    }%
    \hfill
    \subfigure[\ScaleDoc]{
        \includegraphics[
            height=\confidencefigheight,
        ]{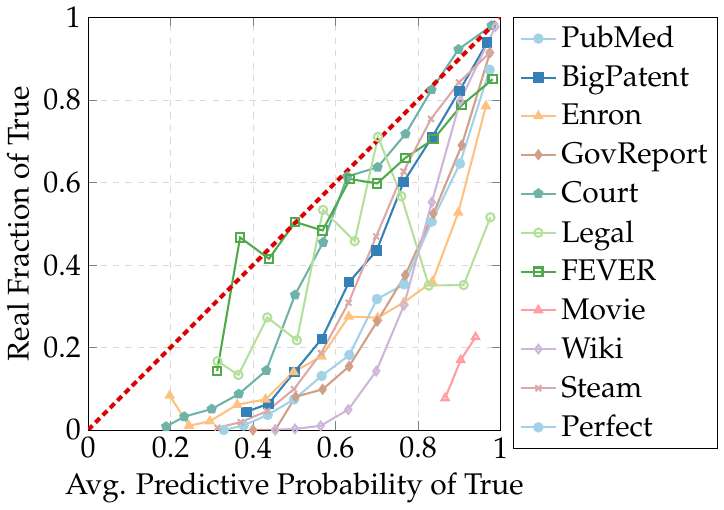}
    }

    \caption{Calibration of proxy confidence (Gemma family)}
    \label{fig:exp:confidence:gemma}
\end{figure}

\begin{figure}[htp]
    \centering
    
    \subfigure[F1 of \Wiki] {
    \includegraphics[width=0.46\linewidth]{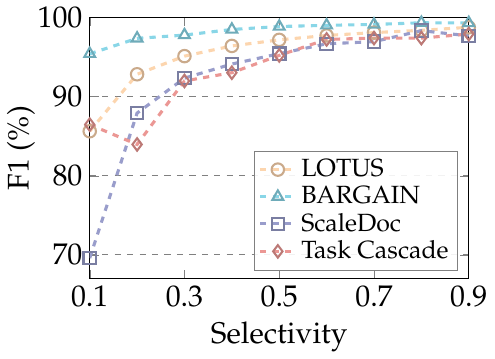}
    }
    \hspace{-3ex} \hfill
    \subfigure[F1 of \Steam] {
    \includegraphics[width=0.46\linewidth]{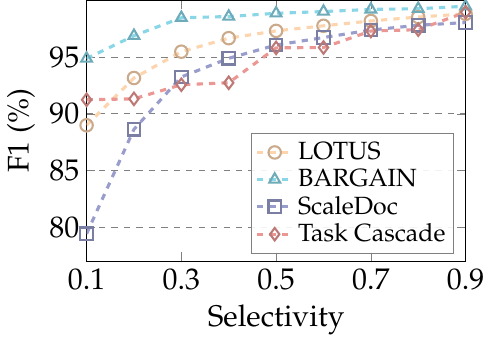}
    }
    \\
    \subfigure[Accuracy of \Wiki] {
    \includegraphics[width=0.46\linewidth]{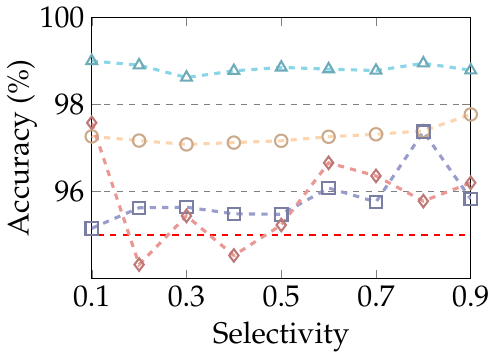}
    }
    \hspace{-3ex} \hfill
    \subfigure[Accuracy of \Steam] {
    \includegraphics[width=0.46\linewidth]{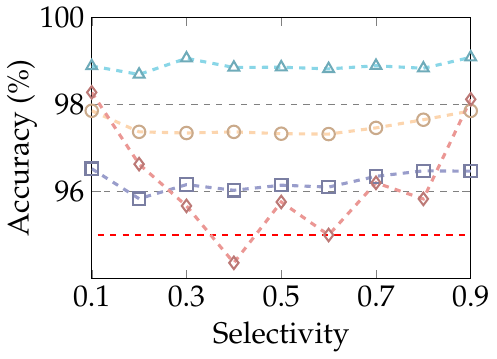}
    }
    \\
     \subfigure[ORR and $\kappa_{0.95}$ of \Wiki] {
    \includegraphics[width=0.46\linewidth]{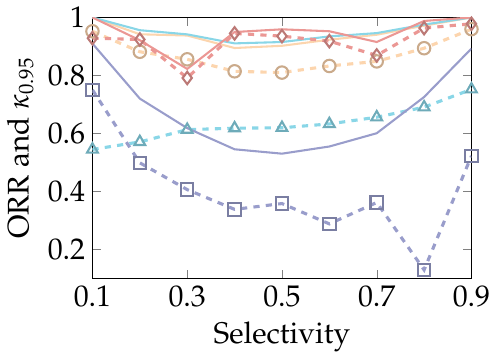}
    }
    \hspace{-3ex} \hfill
    \subfigure[ORR and $\kappa_{0.95}$ of \Steam] {
    \includegraphics[width=0.46\linewidth]{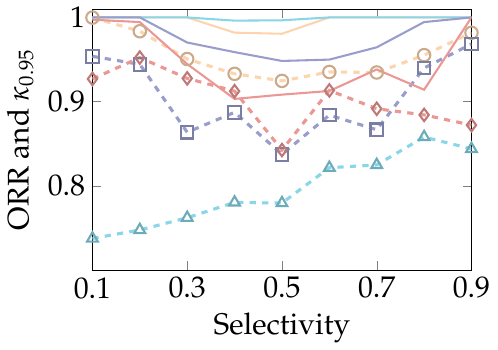}
    }
    
    \caption{Sensitivity to query selectivities (Gemma family)}
    \label{fig:exp:workload_selectivity:Gemma}
\end{figure}

\begin{figure}[htp]
    \centering
     \includegraphics[width=0.88\linewidth]{figures/RQ7/legend.pdf}
    \hspace{-3ex}
    \subfigure[F1 of \Wiki] {
    \includegraphics[width=0.46\linewidth]{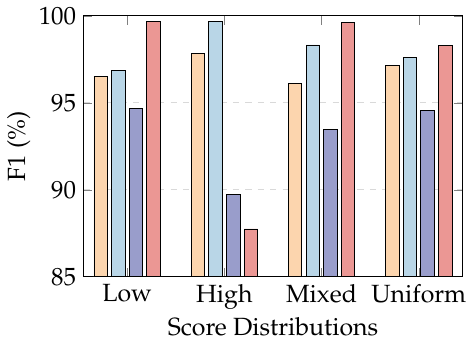}
    }
    \hspace{-3ex} \hfill
    \subfigure[F1 of \Steam] {
    \includegraphics[width=0.46\linewidth]{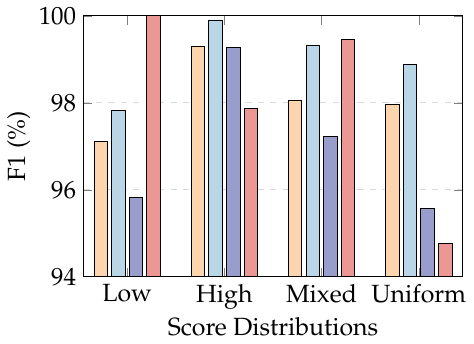}
    }
    \\
    \subfigure[Accuracy of \Wiki] {
    \includegraphics[width=0.46\linewidth]{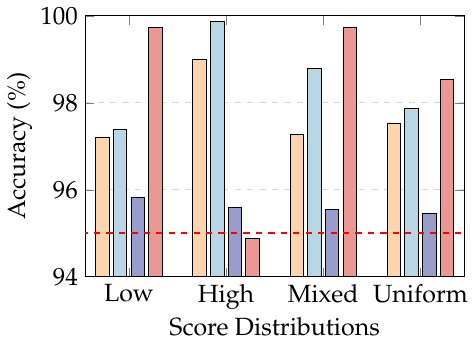}
    }
    \hspace{-3ex} \hfill
    \subfigure[Accuracy of \Steam] {
    \includegraphics[width=0.46\linewidth]{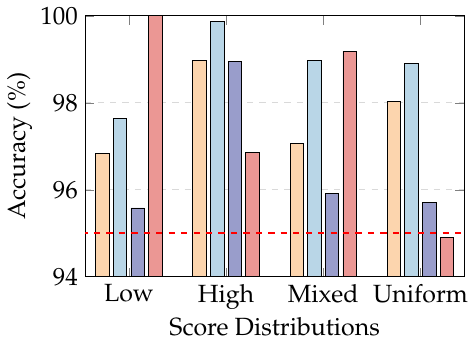}
    }
    \\
     \subfigure[ORR and $\kappa_{0.95}$ of \Wiki] {
    \includegraphics[width=0.46\linewidth]{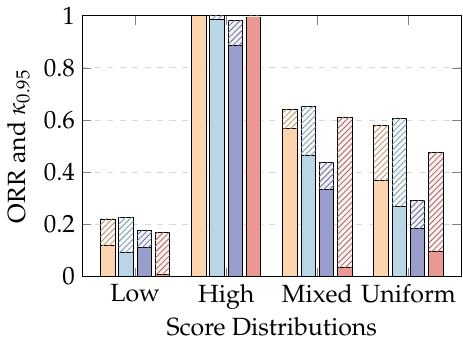}
    }
    \hspace{-3ex} \hfill
    \subfigure[ORR and $\kappa_{0.95}$ of \Steam] {
    \includegraphics[width=0.46\linewidth]{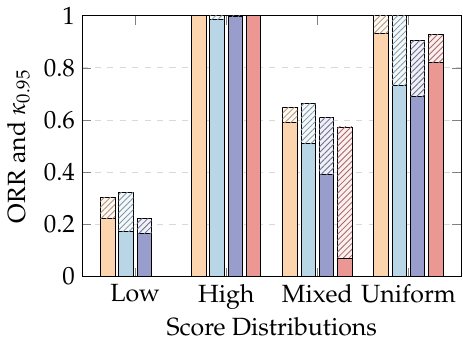}
    }
    
    \caption{Sensitivity by varying oracle confidence score distributions (Gemma family)}
    \label{fig:exp:workload_score:Gemma}
\end{figure}

\begin{figure}
    \centering
    
    \includegraphics[width=\linewidth]{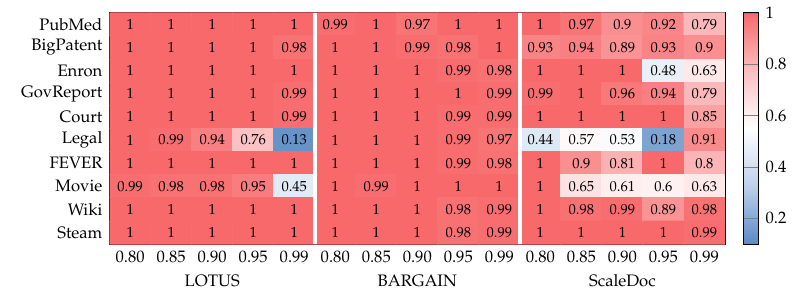}
    \caption{Target Satisfaction Rate (TSR) (Gemma family) }
    \label{fig:exp:TSR:heatmap-gemma}
\end{figure}

\clearpage

\begin{table*}[t]
\centering
\caption{%
Accuracy, ORR, and TSR of the four systems and proxy-switch ablations at  $\alpha=0.95$ (Gemma family). The \colorbox{Best}{best} and \colorbox{SecondBest}{second best} results among the four systems, the \colorbox{Improve}{improved} vs. both components, and the \colorbox{Fail}{unsatisfied} are highlighted. 
}
\label{tab:gemma-overall}
\footnotesize
\setlength{\tabcolsep}{3.5pt}
\begin{tabular}{@{}lccc ccc cc ccc ccc ccc@{}}
\toprule
\multirow{2}{*}{\textbf{Dataset}}
& \multicolumn{3}{c}{\LOTUS}
& \multicolumn{3}{c}{\BARGAIN}
& \multicolumn{2}{c}{\TaskCascade}
& \multicolumn{3}{c}{\ScaleDoc}
& \multicolumn{3}{c}{\makecell[c]{\ScaleDoc proxy \& \\ \LOTUS calibration}}
& \multicolumn{3}{c}{\makecell[c]{\ScaleDoc proxy \& \\ \BARGAIN calibration}} \\
\cmidrule(lr){2-4}
\cmidrule(lr){5-7}
\cmidrule(lr){8-9}
\cmidrule(lr){10-12}
\cmidrule(lr){13-15}
\cmidrule(l){16-18}
& \textbf{Acc.} & \textbf{ORR} & \textbf{TSR}
& \textbf{Acc.} & \textbf{ORR} & \textbf{TSR}
& \textbf{Acc.} & \textbf{ORR}
& \textbf{Acc.} & \textbf{ORR} & \textbf{TSR}
& \textbf{Acc.} & \textbf{ORR} & \textbf{TSR}
& \textbf{Acc.} & \textbf{ORR} & \textbf{TSR} \\
\midrule
\PubMed
& \cellcolor{SecondBest}{97.26} & \cellcolor{SecondBest}{66.63} & \cellcolor{Best}{100/100}
& \cellcolor{Best}{98.39} & 36.56 & \cellcolor{Best}{100/100}
& \cellcolor{Fail}{94.95} & \cellcolor{Best}{74.58}
& 95.68 & 53.92 & \cellcolor{Fail}{92/100}
& \cellcolor{Improve}{97.34} & 41.10 & 100/100
& 97.51 & 17.51 & 98/100 \\

\BigPatent
& \cellcolor{SecondBest}{97.20} & \cellcolor{SecondBest}{78.77} & \cellcolor{Best}{100/100}
& \cellcolor{Best}{98.92} & 58.24 & \cellcolor{SecondBest}{98/100}
& \cellcolor{Fail}{93.97} & \cellcolor{Best}{97.48}
& 95.71 & 40.59 & \cellcolor{Fail}{93/100}
& \cellcolor{Improve}{97.25} & 28.96 & 100/100
& 97.56 & 15.48 & 97/100 \\

\Enron
& \cellcolor{SecondBest}{97.16} & 91.36 & \cellcolor{Best}{100/100}
& \cellcolor{Best}{98.86} & 62.26 & \cellcolor{SecondBest}{99/100}
& \cellcolor{Fail}{94.21} & \cellcolor{Best}{99.87}
& 95.02 & \cellcolor{SecondBest}{97.11} & \cellcolor{Fail}{48/100}
& \cellcolor{Improve}{97.45} & 80.93 & 100/100
& 97.73 & 24.64 & 99/100 \\

\GovReport
& 97.57 & \cellcolor{Best}{88.54} & \cellcolor{Best}{100/100}
& \cellcolor{Best}{98.98} & \cellcolor{SecondBest}{67.32} & \cellcolor{Best}{100/100}
& \cellcolor{SecondBest}{98.17} & 25.90
& 95.93 & 42.62 & \cellcolor{Fail}{94/100}
& 96.94 & 48.88 & 100/100
& 98.14 & 34.89 & 98/100 \\

\Court
& 97.16 & \cellcolor{Best}{89.70} & \cellcolor{Best}{100/100}
& \cellcolor{Best}{98.99} & \cellcolor{SecondBest}{70.02} & \cellcolor{SecondBest}{99/100}
& \cellcolor{SecondBest}{97.99} & 31.78
& 96.24 & 53.80 & \cellcolor{Best}{100/100}
& 97.00 & 52.01 & 100/100
& 98.10 & 38.19 & 100/100 \\

\Legal
& 96.52 & \cellcolor{Best}{70.15} & \cellcolor{Fail}{76/100}
& \cellcolor{Best}{98.70} & \cellcolor{SecondBest}{41.50} & \cellcolor{Best}{99/100}
& \cellcolor{SecondBest}{96.93} & 22.70
& \cellcolor{Fail}{93.54} & 26.25 & \cellcolor{Fail}{18/100}
& 95.62 & 20.58 & \cellcolor{Fail}{69/100}
& \cellcolor{Improve}{99.86} & 0.26 & \cellcolor{Improve}{100/100} \\

\FEVER
& 97.10 & \cellcolor{Best}{89.08} & \cellcolor{Best}{100/100}
& \cellcolor{SecondBest}{98.68} & \cellcolor{SecondBest}{66.11} & \cellcolor{SecondBest}{99/100}
& \cellcolor{Best}{98.70} & 34.18
& 96.35 & 6.58 & \cellcolor{Best}{100/100}
& 96.62 & 26.52 & 100/100
& 97.67 & 19.78 & 99/100 \\

\Movie
& \cellcolor{SecondBest}{97.75} & 26.04 & \cellcolor{SecondBest}{95/100}
& \cellcolor{Best}{98.65} & 2.60 & \cellcolor{Best}{100/100}
& \cellcolor{Fail}{87.20} & \cellcolor{Best}{80.00}
& 95.12 & \cellcolor{SecondBest}{31.61} & \cellcolor{Fail}{60/100}
& 97.11 & 15.87 &\cellcolor{Fail}{86/100}
& \cellcolor{Improve}{99.72} & 0.33 & 100/100 \\

\Wiki
& \cellcolor{SecondBest}{96.13} & \cellcolor{SecondBest}{91.72} & \cellcolor{Best}{100/100}
& \cellcolor{Best}{98.82} & 58.61 & \cellcolor{SecondBest}{98/100}
& \cellcolor{Fail}{94.82} & \cellcolor{Best}{98.13}
& 95.40 & 66.32 & \cellcolor{Fail}{89/100}
& 96.12 & 62.20 & 100/100
& 97.44 & 11.95 & \cellcolor{Improve}{99/100} \\

\Steam
& \cellcolor{SecondBest}{96.21} & \cellcolor{Best}{97.62} & \cellcolor{Best}{100/100}
& \cellcolor{Best}{98.81} & 79.91 & \cellcolor{SecondBest}{98/100}
& 95.04 & 87.61
& 96.05 & \cellcolor{SecondBest}{91.69} & \cellcolor{Best}{100/100}
& 96.16 & 92.14 & 100/100
& 98.29 & 37.12 & 98/100 \\

\midrule
\textbf{Overall}
& \cellcolor{SecondBest}{97.01} & \cellcolor{Best}{78.96} & \cellcolor{SecondBest}{971/1000}
& \cellcolor{Best}{98.78} & 54.31 & \cellcolor{Best}{990/1000}
& 95.20 & \cellcolor{SecondBest}{65.22}
& 95.50 & 51.05 & \cellcolor{Fail}{794/1000}
& 96.76 & 46.92 & 955/1000
& 98.20 & 20.02 & 988/1000 \\
\bottomrule
\end{tabular}%
\end{table*}

\begin{figure*}[tp]
        \includegraphics[width=0.35\linewidth]{figures/RQ5/kappa_ORR-Qwen/legend.pdf}
    \par\vspace{0.5em}
    \subfigure[\PubMed] {
    \includegraphics[width=0.20\linewidth]{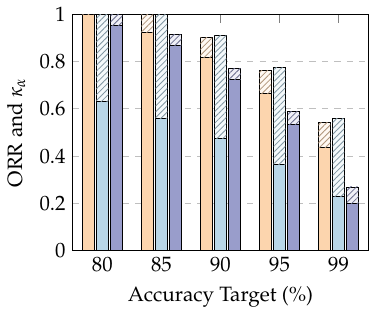}
    } 
    \hspace{-2ex}
    \subfigure[\BigPatent] {
    \includegraphics[width=0.18\linewidth]{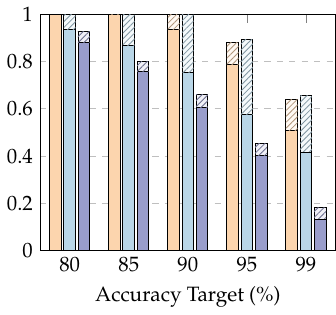}
    }
    \hspace{-2ex}
    \subfigure[\Enron] {
    \includegraphics[width=0.18\linewidth]{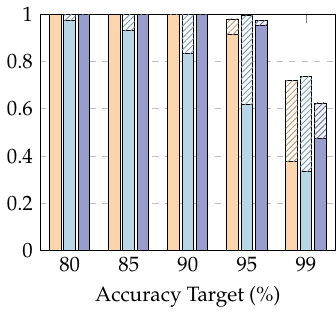}
    }
    \hspace{-2ex}
    \subfigure[\GovReport] {
    \includegraphics[width=0.18\linewidth]{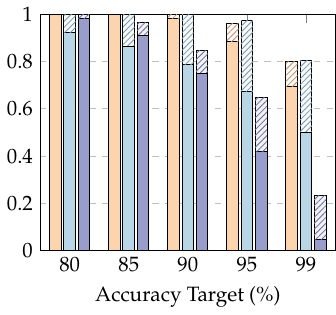}
    }
    \hspace{-2ex}
    \subfigure[\Court] {
    \includegraphics[width=0.18\linewidth]{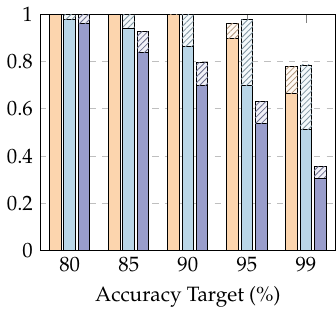}
    }
    \\
    
     \subfigure[\Legal] {
    \includegraphics[width=0.20\linewidth]{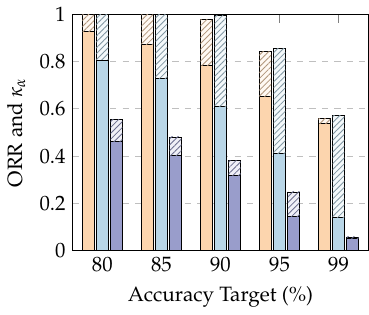}
    }
    \hspace{-2ex}
    \subfigure[\FEVER] {
    \includegraphics[width=0.18\linewidth]{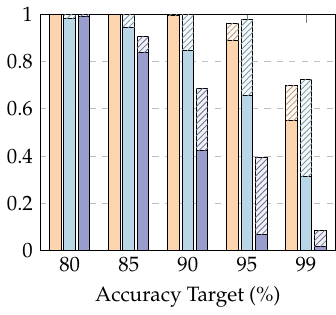}
    }
    \hspace{-2ex}
    \subfigure[\Movie] {
    \includegraphics[width=0.18\linewidth]{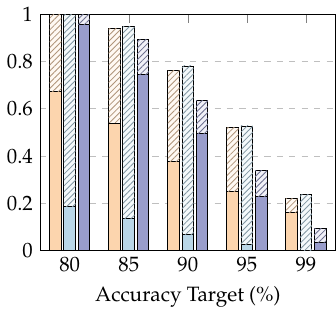}
    }
    \hspace{-2ex}
    \subfigure[\Wiki] {
    \includegraphics[width=0.18\linewidth]{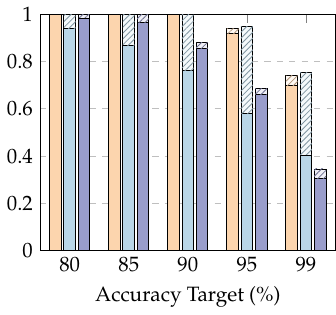}
    }
    \hspace{-2ex}
    \subfigure[\Steam] {
    \includegraphics[width=0.18\linewidth]{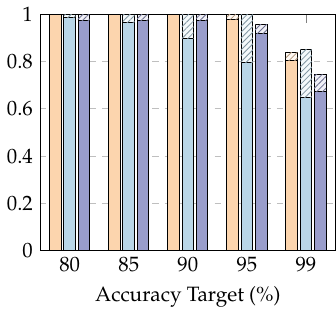}
    }

    \caption{Oracle reduction rate and $\kappa_\alpha$ under by varying accuracy target $\alpha$ (Gemma family)}
    \label{fig:exp:ORR_kappa:Gemma}
\end{figure*}

\begin{figure*}[tp]
    \subfigure[\PubMed] {
    \includegraphics[width=0.195\linewidth]{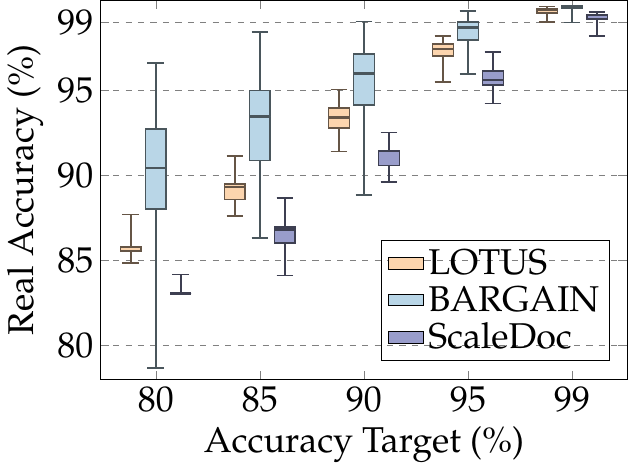}
    } 
    \hspace{-2ex}
    \subfigure[\BigPatent] {
    \includegraphics[width=0.18\linewidth]{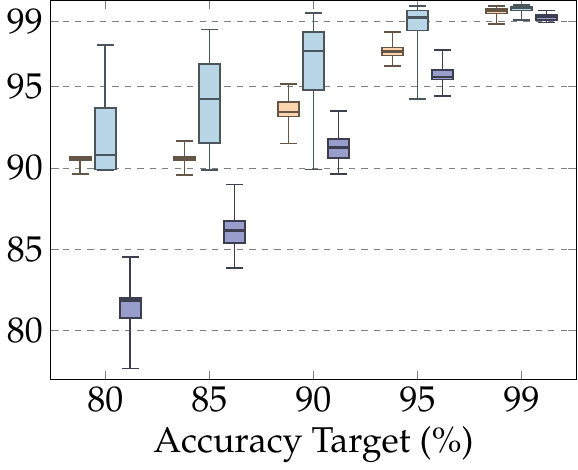}
    }
    \hspace{-2ex}
    \subfigure[\Enron] {
    \includegraphics[width=0.18\linewidth]{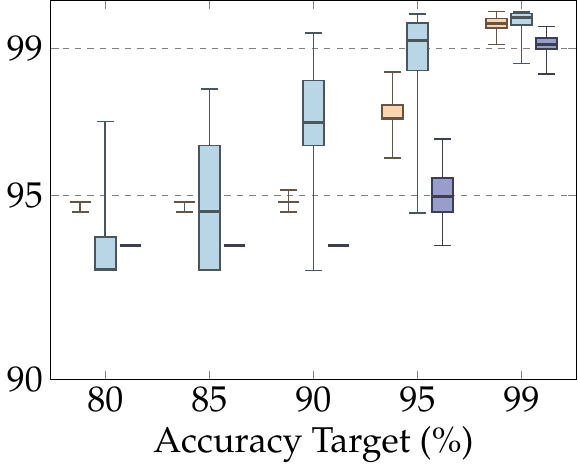}
    }
    \hspace{-2ex}
    \subfigure[\GovReport] {
    \includegraphics[width=0.18\linewidth]{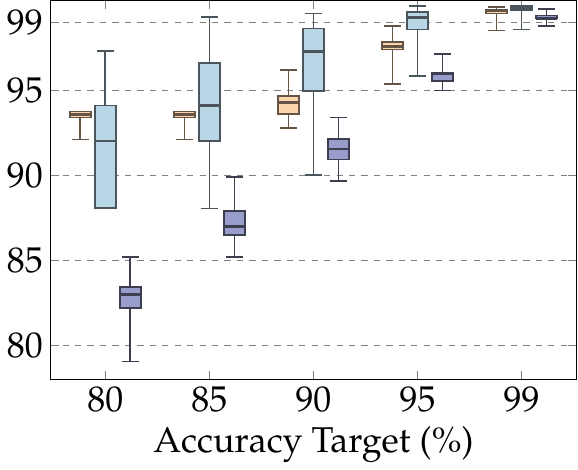}
    }
    \hspace{-2ex}
    \subfigure[\Court] {
    \includegraphics[width=0.18\linewidth]{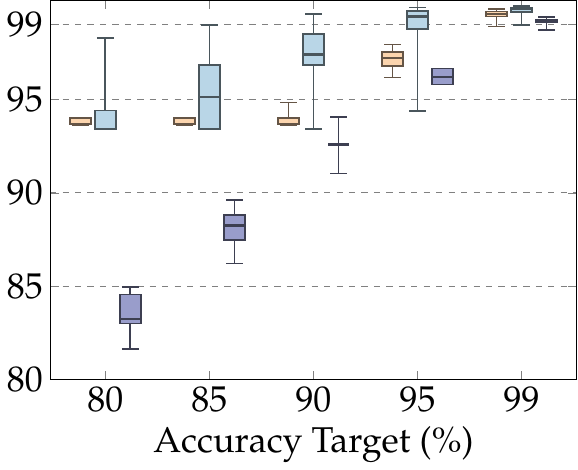}
    }
    \\
    
     \subfigure[\Legal] {
    \includegraphics[width=0.195\linewidth]{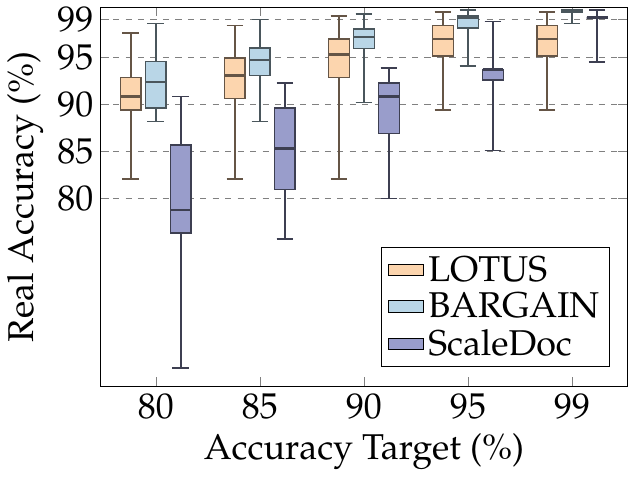}
    }
    \hspace{-2ex}
    \subfigure[\FEVER] {
    \includegraphics[width=0.18\linewidth]{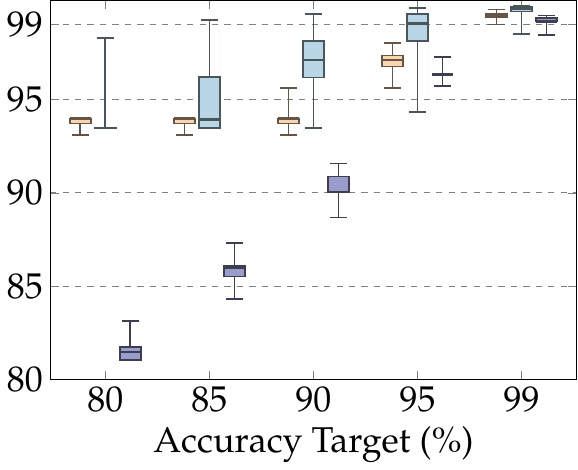}
    }
    \hspace{-2ex}
    \subfigure[\Movie] {
    \includegraphics[width=0.18\linewidth]{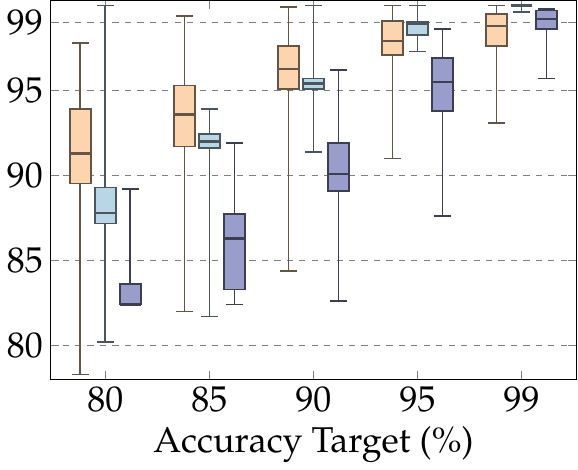}
    }
     \hspace{-2ex}
    \subfigure[\Wiki] {
    \includegraphics[width=0.18\linewidth]{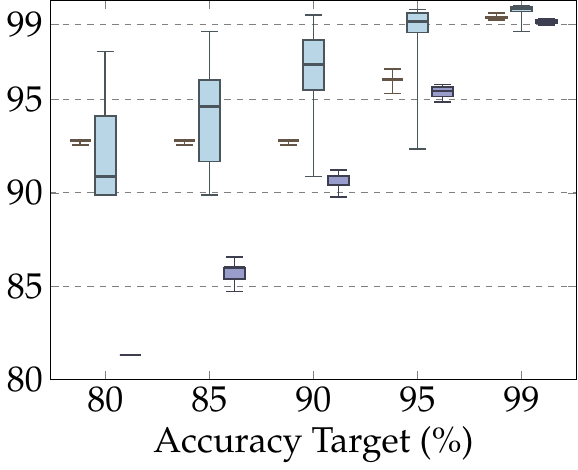}
    }
    \hspace{-2ex}
    \subfigure[\Steam] {
    \includegraphics[width=0.18\linewidth]{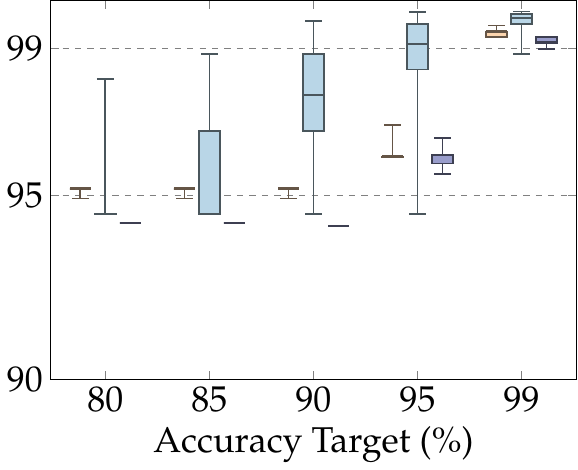}
    }

    \caption{Accuracy as the target $\alpha$ varies(Gemma family)}
    \label{fig:exp:accuracy:Gemma}
\end{figure*}

\begin{figure*}[tp]
    \subfigure[\PubMed] {
    \includegraphics[width=0.195\linewidth]{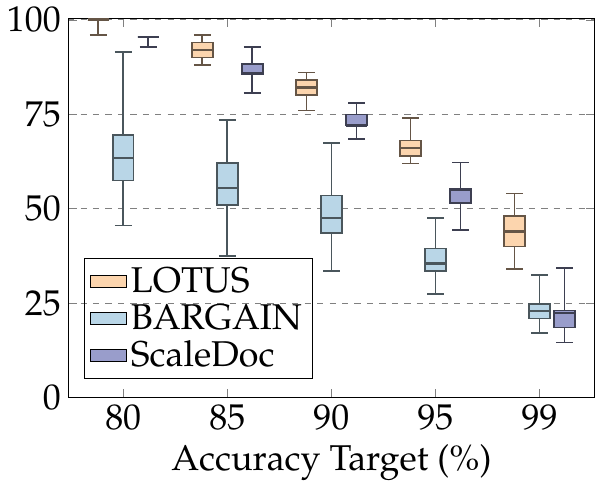}
    } 
    \hspace{-2ex}
    \subfigure[\BigPatent] {
    \includegraphics[width=0.18\linewidth]{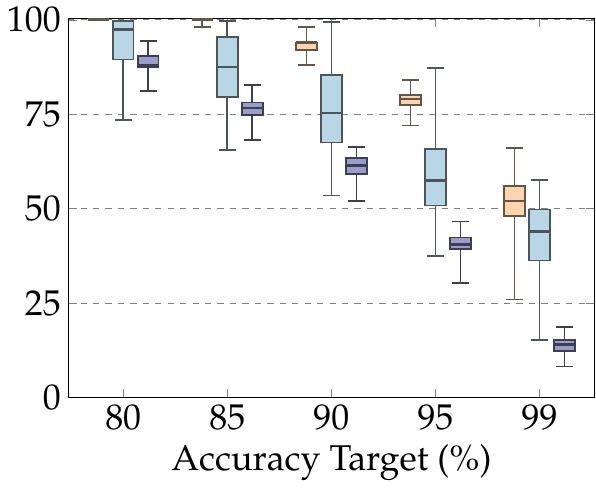}
    }
    \hspace{-2ex}
    \subfigure[\Enron] {
    \includegraphics[width=0.18\linewidth]{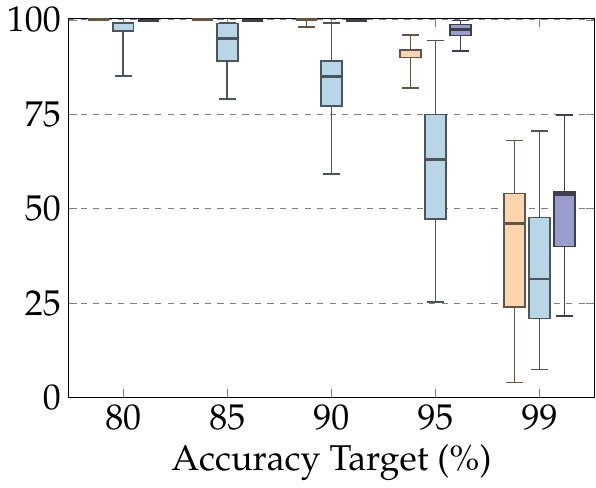}
    }
    \hspace{-2ex}
    \subfigure[\GovReport] {
    \includegraphics[width=0.18\linewidth]{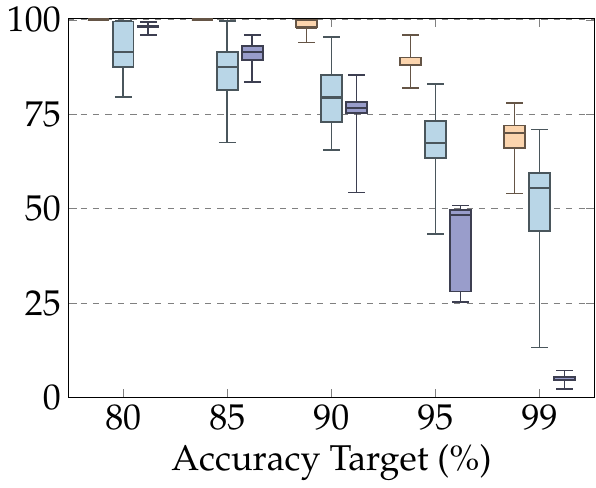}
    }
    \hspace{-2ex}
    \subfigure[\Court] {
    \includegraphics[width=0.18\linewidth]{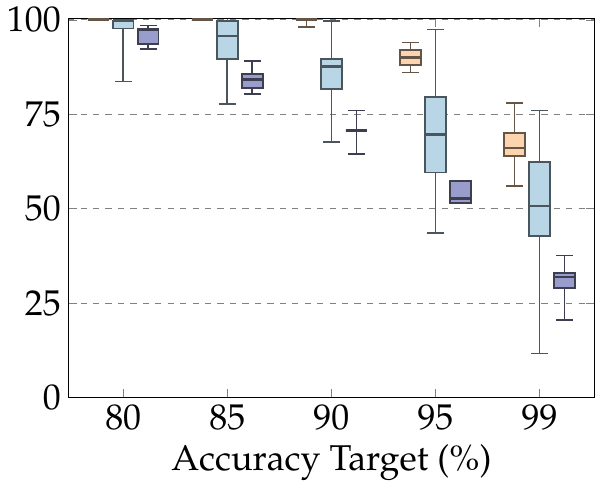}
    }
    \\
    
     \subfigure[\Legal] {
    \includegraphics[width=0.195\linewidth]{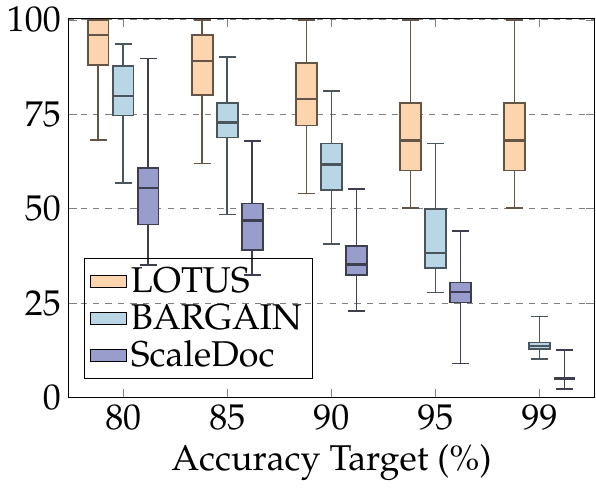}
    }
    \hspace{-2ex}
    \subfigure[\FEVER] {
    \includegraphics[width=0.18\linewidth]{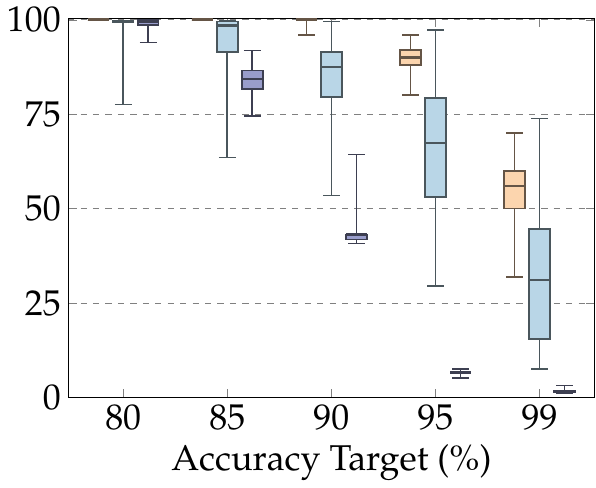}
    }
    \hspace{-2ex}
    \subfigure[\Movie] {
    \includegraphics[width=0.18\linewidth]{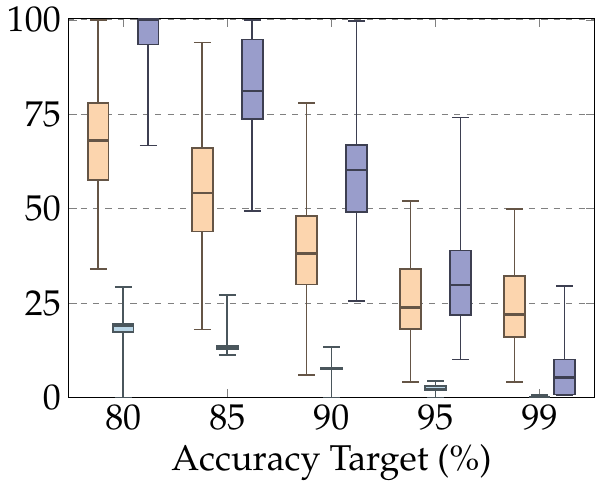}
    }
     \hspace{-2ex}
    \subfigure[\Wiki] {
    \includegraphics[width=0.18\linewidth]{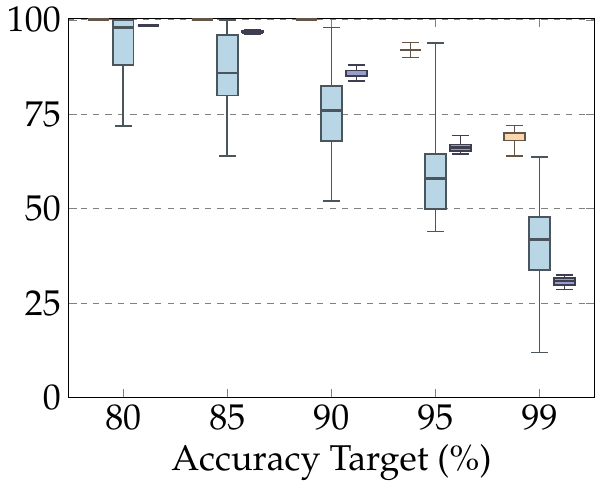}
    }
    \hspace{-2ex}
    \subfigure[\Steam] {
    \includegraphics[width=0.18\linewidth]{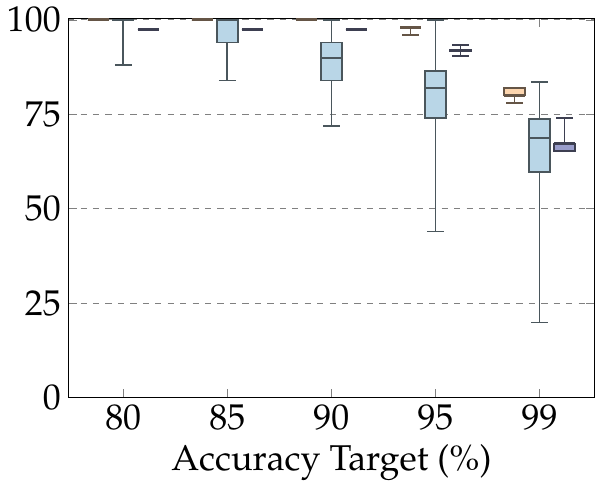}
    }

    \caption{ORR as the target $\alpha$ varies(Gemma family)}
    \label{fig:exp:accuracy:Gemma}
\end{figure*}

\begin{figure*}[tp]
    \subfigure[\PubMed] {
    \includegraphics[width=0.195\linewidth]{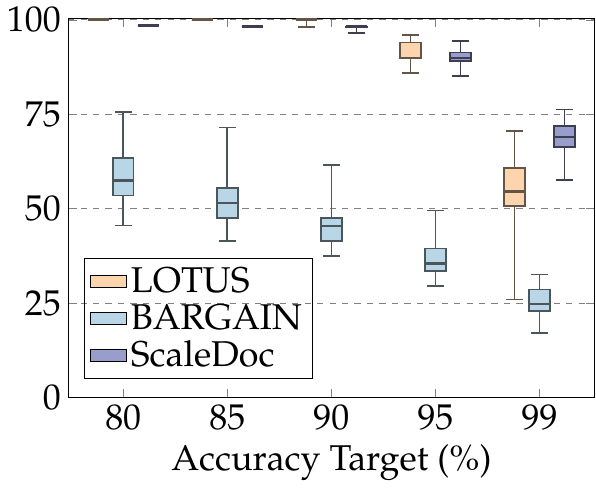}
    } 
    \hspace{-2ex}
    \subfigure[\BigPatent] {
    \includegraphics[width=0.18\linewidth]{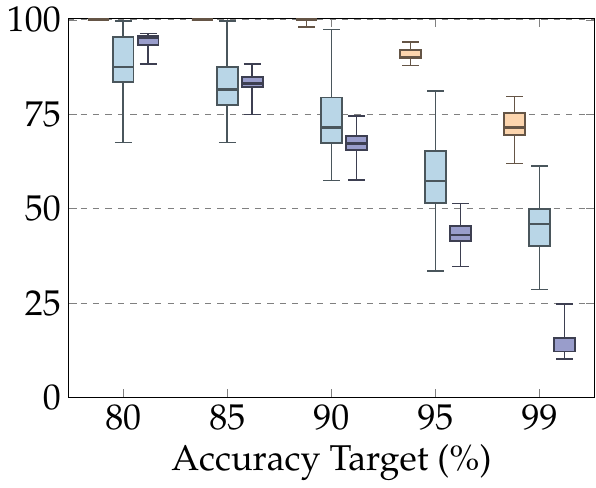}
    }
    \hspace{-2ex}
    \subfigure[\Enron] {
    \includegraphics[width=0.18\linewidth]{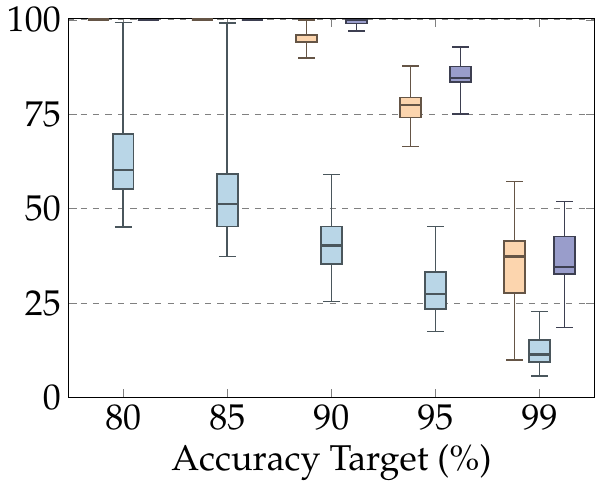}
    }
    \hspace{-2ex}
    \subfigure[\GovReport] {
    \includegraphics[width=0.18\linewidth]{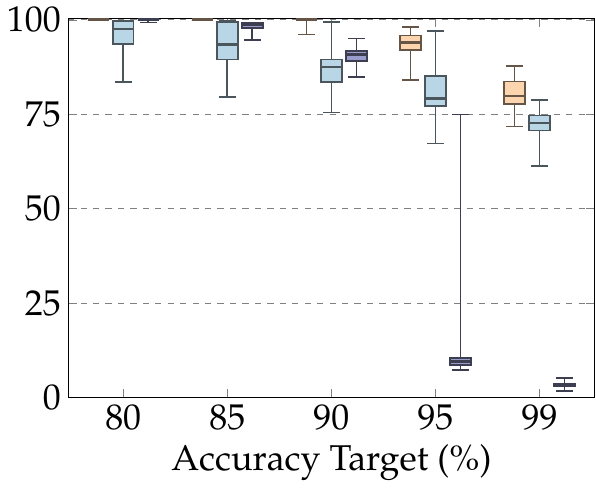}
    }
    \hspace{-2ex}
    \subfigure[\Court] {
    \includegraphics[width=0.18\linewidth]{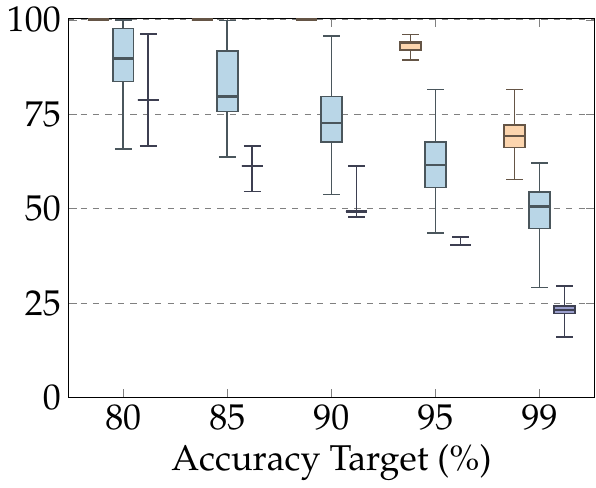}
    }
    \\
    
     \subfigure[\Legal] {
    \includegraphics[width=0.195\linewidth]{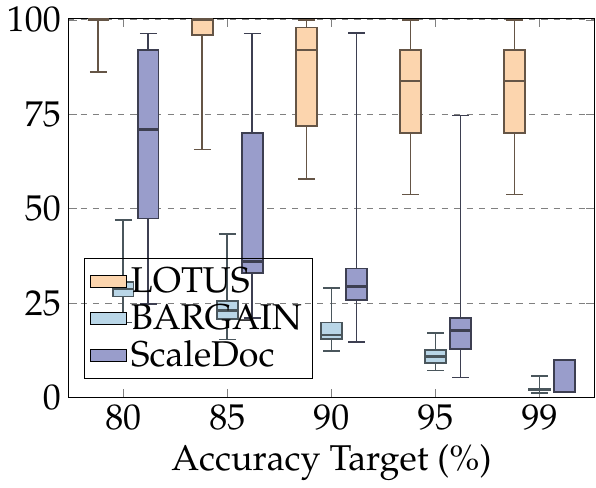}
    }
    \hspace{-2ex}
    \subfigure[\FEVER] {
    \includegraphics[width=0.18\linewidth]{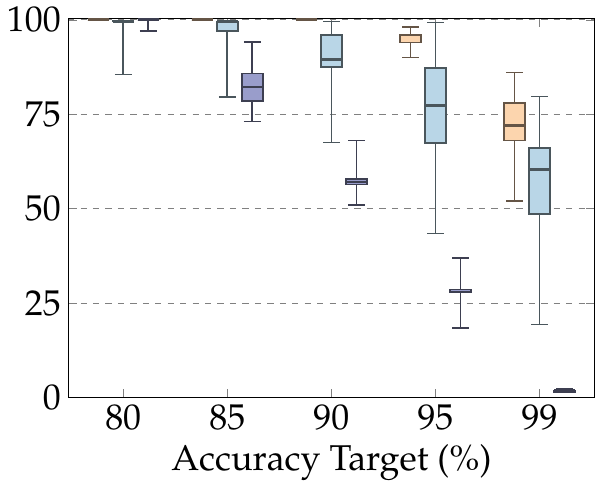}
    }
    \hspace{-2ex}
    \subfigure[\Movie] {
    \includegraphics[width=0.18\linewidth]{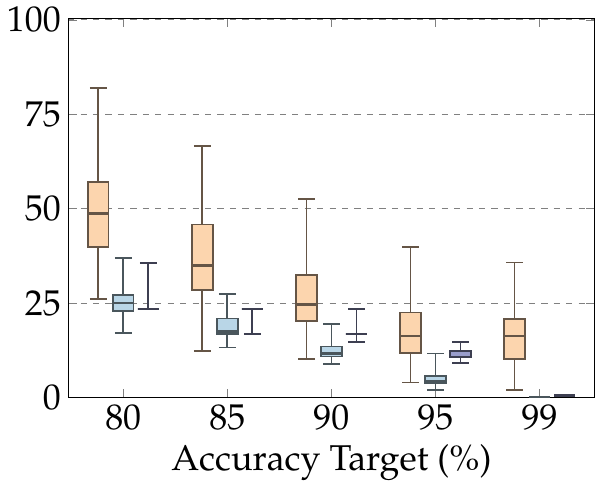}
    }
     \hspace{-2ex}
    \subfigure[\Wiki] {
    \includegraphics[width=0.18\linewidth]{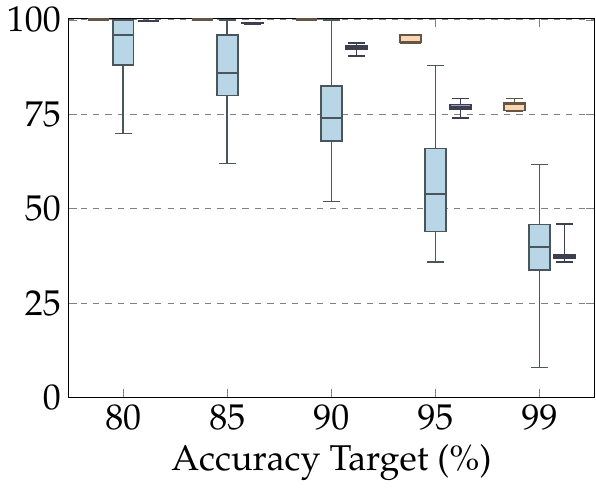}
    }
    \hspace{-2ex}
    \subfigure[\Steam] {
    \includegraphics[width=0.18\linewidth]{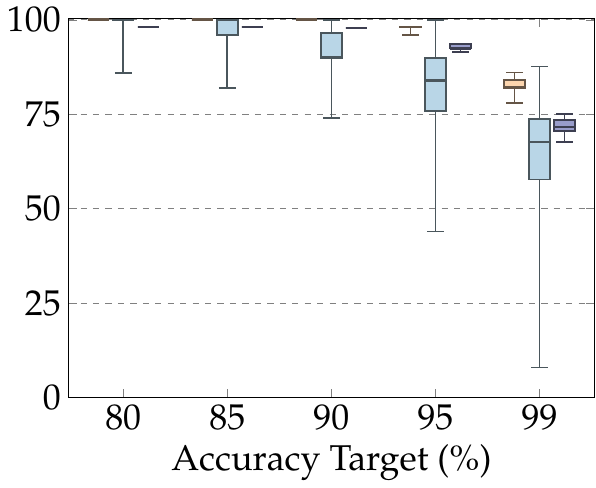}
    }

    \caption{ORR as the target $\alpha$ varies(Qwen family)}
    \label{fig:exp:accuracy:Gemma}
\end{figure*}

\begin{figure*}[tp]
    \centering
    \includegraphics[width=0.55\linewidth]{%
        figures/RQ4/cost/appendix_a_s7_legend.pdf}
    \par\vspace{0.35em}
    \subfigure[\PubMed]{%
        \includegraphics[width=0.25\linewidth]{%
            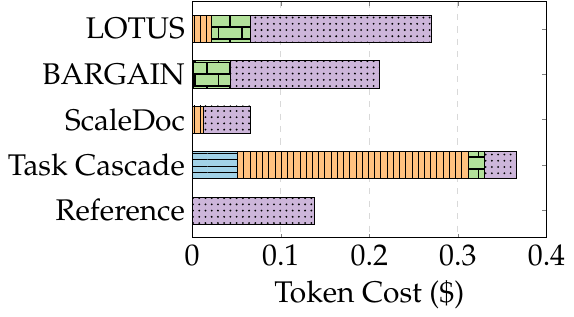}%
    }%
    \hfill
    \subfigure[\BigPatent]{%
        \includegraphics[width=0.17\linewidth]{%
            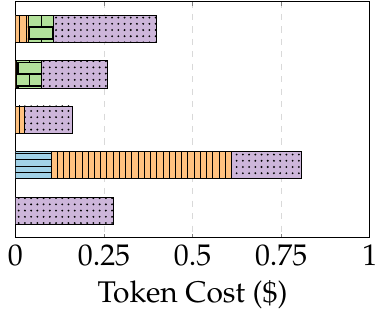}%
    }%
    \hfill
    \subfigure[\Enron]{%
        \includegraphics[width=0.17\linewidth]{%
            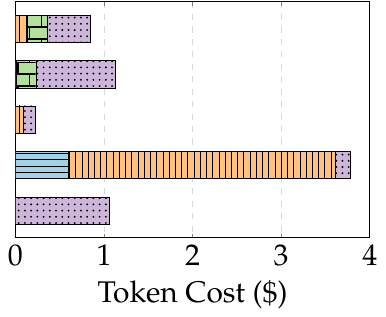}%
    }%
    \hfill
    \subfigure[\GovReport]{%
        \includegraphics[width=0.17\linewidth]{%
            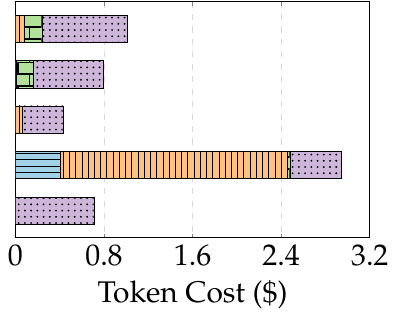}%
    }%
    \hfill
    \subfigure[\Court]{%
        \includegraphics[width=0.17\linewidth]{%
            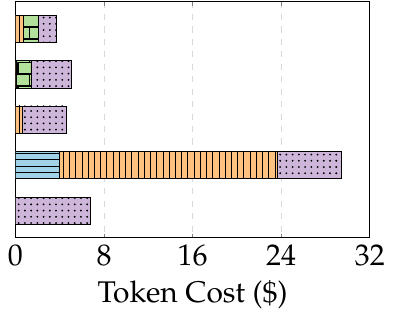}%
    }%
    \par\vspace{0.3em}
    \subfigure[\Legal]{%
        \includegraphics[width=0.25\linewidth]{%
            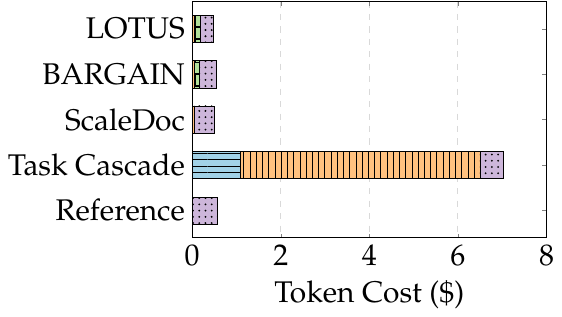}%
    }%
    \hfill
    \subfigure[\FEVER]{%
        \includegraphics[width=0.17\linewidth]{%
            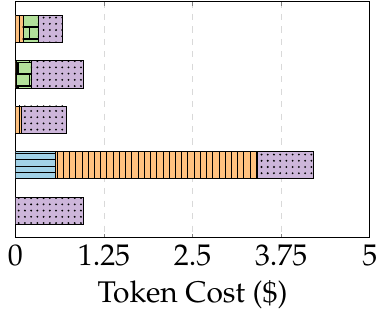}%
    }%
    \hfill
    \subfigure[\Movie]{%
        \includegraphics[width=0.17\linewidth]{%
            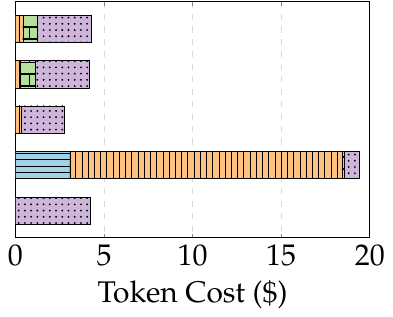}%
    }%
    \hfill
    \subfigure[\Wiki]{%
        \includegraphics[width=0.17\linewidth]{%
            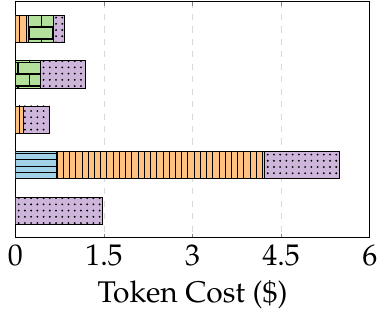}%
    }%
    \hfill
    \subfigure[\Steam]{%
        \includegraphics[width=0.17\linewidth]{%
            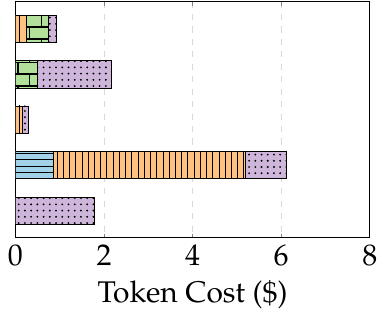}%
    }%
    
    \par\vspace{0.2em}
    \caption{Token consumption (\$) (Gemma oracle)}
    \label{fig:exp:token_cost:Gemma}
\end{figure*}

\begin{figure*}[tp]
    \centering
    \includegraphics[width=0.35\linewidth]{%
        figures/RQ4/latency/appendix_a_s7_legend.pdf}
    \par\vspace{0.35em}
    \subfigure[\PubMed]{%
        \includegraphics[width=0.25\linewidth]{%
            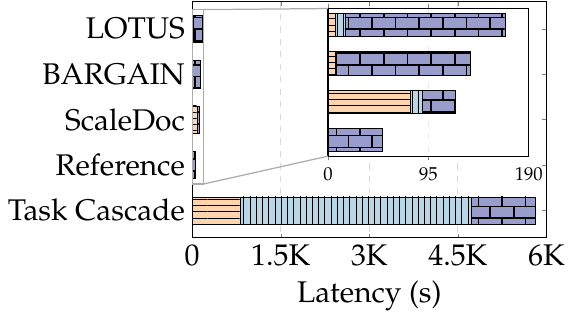}%
    }%
    \hfill
    \subfigure[\BigPatent]{%
        \includegraphics[width=0.17\linewidth]{%
            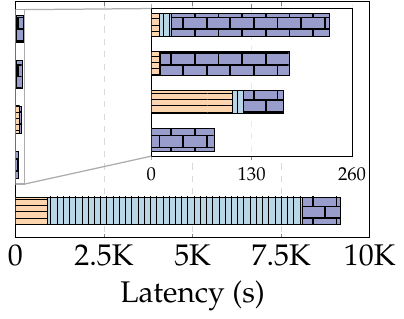}%
    }%
    \hfill
    \subfigure[\Enron]{%
        \includegraphics[width=0.17\linewidth]{%
            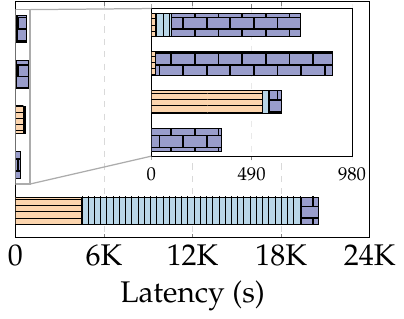}%
    }%
    \hfill
    \subfigure[\GovReport]{%
        \includegraphics[width=0.17\linewidth]{%
            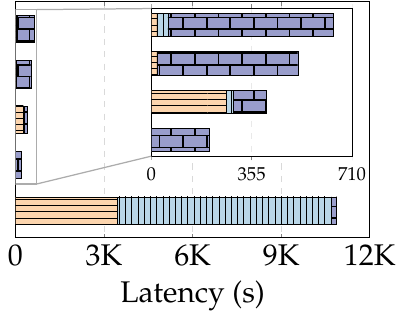}%
    }%
    \hfill
    \subfigure[\Court]{%
        \includegraphics[width=0.17\linewidth]{%
            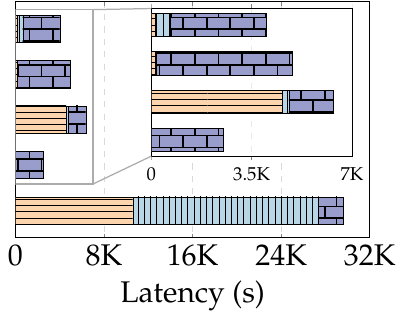}%
    }%
    \par\vspace{0.3em}
    \subfigure[\Legal]{%
        \includegraphics[width=0.25\linewidth]{%
            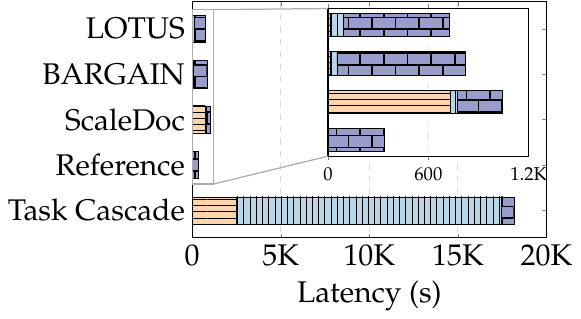}%
    }%
    \hfill
    \subfigure[\FEVER]{%
        \includegraphics[width=0.17\linewidth]{%
            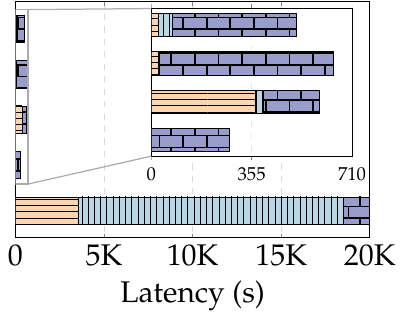}%
    }%
    \hfill
    \subfigure[\Movie]{%
        \includegraphics[width=0.17\linewidth]{%
            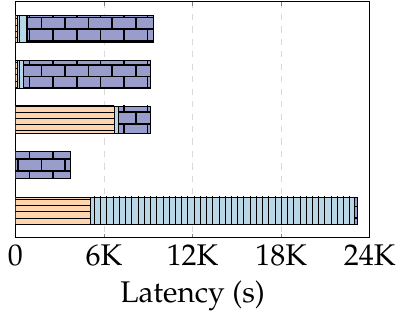}%
    }%
    \hfill
    \subfigure[\Wiki]{%
        \includegraphics[width=0.17\linewidth]{%
            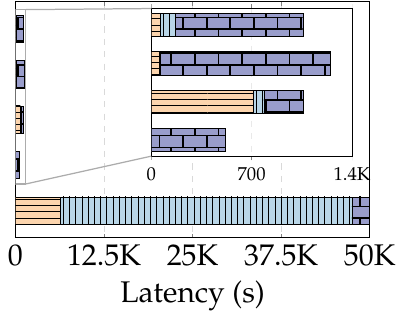}%
    }%
    \hfill
    \subfigure[\Steam]{%
        \includegraphics[width=0.17\linewidth]{%
            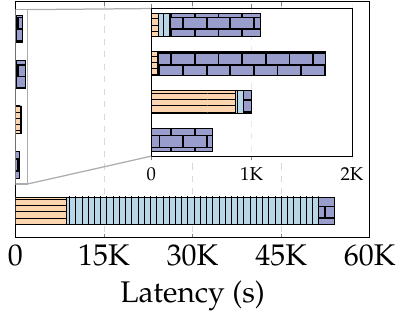}%
    }%
    
    \par\vspace{0.2em}
    \caption{End-to-end latency (seconds) (Gemma family)}
    \label{fig:exp:latency:Gemma}
\end{figure*}

\end{document}